\pdfoutput=1
\documentclass{emulateapj}
\usepackage{color}
\usepackage{amsmath}
\usepackage{float}
\usepackage{stmaryrd}

\newcommand{\etal}{{et al.\/}}

\shorttitle{Faraday Conversion in Turbulent Blazar Jets}
\shortauthors{MacDonald \& Marscher}

\usepackage{mathtools}
\usepackage[x11names]{xcolor}
\newtagform{blue}{\color{RoyalBlue3}(}{)}
\newtagform{redandblue}[\textcolor{RoyalBlue3}]{\color{red}(}{)}

\begin{document}

\title{Faraday Conversion in Turbulent Blazar Jets}

\author{Nicholas R. MacDonald\altaffilmark{1,2} and Alan P. Marscher\altaffilmark{2}}
\affil{$~^{1}$Max-Planck-Institut f\"{u}r Radioastronomie, Auf dem H\"{u}gel 69, D- 53121 Bonn, Germany}
\affil{$~^{2}$Institute for Astrophysical Research, Boston University, 725 Commonwealth Avenue, Boston, MA 02215}

\begin{abstract}
Low ($\lesssim 1\%$) levels of circular polarization (CP) detected at radio frequencies in the relativistic jets of some blazars can provide insight into the underlying nature of the jet plasma.  CP can be produced through linear birefringence, in which initially linearly polarized emission produced in one region of the jet is altered by Faraday rotation as it propagates through other regions of the jet with varying magnetic field orientation. Marscher has begun a study of jets with such magnetic geometries using the Turbulent Extreme Multi-Zone (TEMZ) model, in which turbulent plasma crossing a standing shock in the jet is represented by a collection of thousands of individual plasma cells, each with distinct magnetic field orientations. Here we develop a radiative transfer scheme that allows the numerical TEMZ code to produce simulated images of the time-dependent linearly and circularly polarized intensity at different radio frequencies.  In this initial study, we produce synthetic polarized emission maps that highlight the linear and circular polarization expected within the model.
\end{abstract}

\keywords{blazars: non-thermal radiative transfer - relativistic processes - polarization}

\section{Introduction}

Observations of polarized synchrotron radio emission from the jets of blazars provide a probe of the underlying magnetic field structure that is thought to play an important role in the dynamics of these highly relativistic flows. Low levels (typically $<$ 0.5 \% Stokes I) of circularly polarized intensity (Stokes $V$) have been detected in a sub-set of blazar jets \citep{wardle98,homan99,homan01,homan09}. Circular polarization (CP) can be used to probe the plasma composition of these jets \citep{wardle98}.  Unlike linear polarization (LP), which can be altered by Faraday rotation in magnetized plasma both within and external to the jet, CP is produced at levels $< 1\%$ by the basic synchrotron process, or at potentially higher levels through Faraday conversion from LP to CP as the radiation propagates through the jet.  A measurement of the fractional CP ($m_{\rm c} \equiv -V/I$) emanating from the jet can probe the charge asymmetry of leptons within the jet plasma, and therefore the ratio of the number of positrons to protons \citep{wardle98, ruszkowski02, homan09}.

LP from blazars at high radio frequencies ranges from several percent or less to tens of percent, with the higher values tending to occur downstream of the most compact emission (the ``core'') seen on very long baseline interferometry (VLBI) images \citep[e.g.,][]{cawthorne1993,marscher02}. The LP of blazars at optical wavelengths \citep[e.g.,][]{smith16} and in the millimeter-wave core \citep{jorstad07} is both variable and weaker than the $70$-$75\%$ value from synchrotron emission in a uniform magnetic field, characteristics that can be explained through turbulent disordering of the field. With these observations in mind, \cite{marscher14} has created the Turbulent Extreme Multi-Zone (TEMZ) model for blazar emission. The TEMZ code simulates the non-thermal emission from thousands of turbulent plasma cells moving relativistically across a standing conical shock (Figure 1).  Each cell of plasma within the model contains a distinct magnetic field that combines a turbulent component with an underlying ordered component.  The combined emission from these turbulent cells reproduces (statistically) the LP that is observed in blazars, including the observed level of variability in both degree and position angle \citep{marscher15}.  The turbulent nature of this jet model can potentially create the characteristics of a birefringent plasma in which Faraday conversion can produce circularly polarized emission \citep{ruszkowski02}.

Here we develop a numerical algorithm to carry out polarized radiative transfer through the TEMZ model.  We have embedded this algorithm into the ray-tracing code RADMC-3D (see http://ascl.net/1202.015), which we use to image the TEMZ grid.  This makes it possible to compute both LP and CP intensity images as a function of both time and frequency that include the effects of Faraday rotation as well as LP to CP conversion. We combine the turbulently disordered magnetic field with an ordered component, varying the ratio between the two. The results can be compared to VLBI images as well as to single-dish polarization measurements.

This paper is organized as follows: In \S2 we summarize the TEMZ model, in \S3 we outline the radiative transfer theory adopted in our study, and in \S4 we list the model parameters and present an initial ray-tracing calculation through an ordered magnetic field.  In \S5 we investigate the effect turbulence has on the birefringence of the plasma, while in \S6 we carry out a ray-tracing calculation through this turbulent magnetic field. In \S7 \& \S8 we present synthetic multi-epoch observations of the integrated levels of circular polarization.  In \S9 we explore internal Faraday rotation within the TEMZ model. In \S10 we investigate how CP depends on plasma composition, and in \S11 we study spectral polarimetry.  Finally, in \S12 we present our summary and conclusions. We adopt the following cosmological parameters: $H_{\rm o} = 71 ~ \rm km ~ \rm s^{-1} ~ \rm Mpc^{-1}$, $\Omega_{\rm m} = 0.27$, and $\Omega_{\Lambda} = 0.73$. 

\begin{figure*}[!ht]
  \setlength{\abovecaptionskip}{-6pt}
  \begin{center}
    \scalebox{0.78}{\includegraphics[width=2.0\columnwidth,clip]{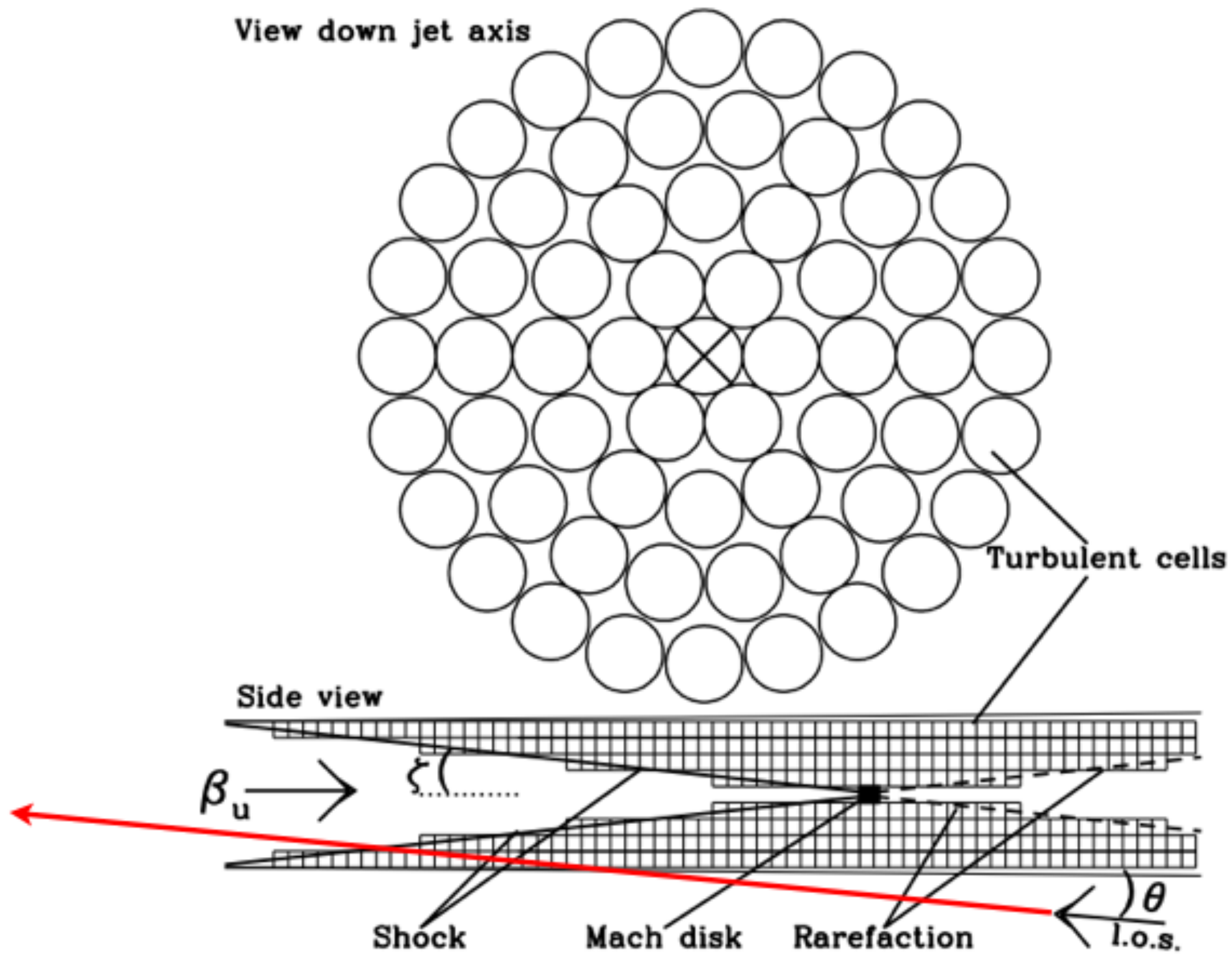}}
  \end{center}
  \caption{\label{fig1}A schematic representation of the TEMZ model, reproduced from \cite{marscher14} and \cite{nick16}.  The radio core of a blazar is modeled as a collection of thousands of turbulent cells of plasma moving relativistically across a standing shock. The calculations presented in this study include 169 cells in the cross-section that includes the Mach disk (a small, transversely oriented shock), and a total of $13,440$ cells along the jet. The red line depicts how, along a given sight-line through the TEMZ grid, an individual ray crosses multiple cells of plasma, each with a distinct magnetic field orientation.}
\end{figure*}

\pagebreak

\section{Summary of the TEMZ Model}

In the TEMZ code \citep{marscher14}, the jet plasma is divided into thousands of cylindrical cells (as depicted in Figure \ref{fig1}), each with uniform physical properties. Simulated turbulence is realized through an adaptation of a scheme by \cite{jones88}, in which each cell belongs to four nested zones (with dimensions of $1\times1\times1$, $2\times2\times2$, $4\times4\times4$, and $8\times8\times8$ cells), with each zone contributing a fraction of the turbulent magnetic field and density of the cells that it contains. These contributions follow a Kolmogorov spectrum to determine the relation between the size of the zone and the fraction of the turbulent field component or electron density (which can include positrons as well) that the zone provides to its cells. An ordered component of the magnetic field can optionally be included, superposed on the turbulent field. The energy density of the particles injected at the upstream end of the jet varies according to a power-law red-noise spectrum. The magnitudes of the turbulent field and density of each zone are computed randomly within a log-normal distribution, while the direction of the field is also determined randomly at the upstream and downstream borders of the zone and rotated smoothly in between. This scheme qualitatively reproduces both the randomness of turbulence and correlations among turbulent cells located near each other. The direction of the turbulent velocity of each cell is chosen randomly (with magnitude set at a specified fraction of the sound speed), and this velocity is added vectorially to the systemic velocity of the jet flow. 

The turbulence is assumed to endow the plasma with relativistic electrons that follow a power-law energy distribution. When a plasma cell crosses a standing oblique (conical in three dimensions) shock in the millimeter-wave core, its electrons are heated solely via compression if the magnetic field is not close to parallel to the shock normal, or accelerated to considerably higher energies (while maintaining the power-law form) if
the field is more favorably oriented \cite[see, e.g.,][]{summerlin12}. The component of the magnetic field lying perpendicular to the shock normal, as well as the density, are amplified by shock compression. 
After passing through the shock front the electron energy distributions within the downstream plasma cells are subjected to synchrotron losses (\citealt{pacholczyk70}):
\begin{equation}\label{eqn0}
\dot{\gamma} = -k_{r}( B^{2} + 8\pi u_{\rm{ph}})\gamma^{2},
\end{equation}
where $k_{r} = 1.3 \times 10^{-9} ~ \rm{erg}^{-1} ~ \rm{s}^{-1} ~ \rm{cm}^{3}$ and $u_{\rm{ph}}$ is the energy density of the photon field within each cell (i.e., inverse-Compton losses). Downstream of the shock, a conical rarefaction decompresses the plasma; beyond this point the emission is considered to contribute only a small fraction of the flux and hence is ignored. TEMZ computes the synchrotron emission and absorption coefficients in each cell as viewed by the observer, taking into account Doppler beaming, the aberrated direction of the magnetic field \citep[see, e.g.,][]{lyutikov05}, and light-travel delays. It also calculates inverse Compton emission and full time-dependent spectral energy distributions.

\newpage

\begin{widetext}
\vspace*{1.0cm}
\begin{equation}\label{eqn1}\small{
\left \{ \begin{array}{cccc}
\left( \dfrac{d}{dl} + \kappa_{I} \right) & \kappa_{Q} & \kappa_{U} & \kappa_{V} \\
\kappa_{Q} & \left( \dfrac{d}{dl} + \kappa_{I} \right) & \kappa^{*}_{~ V} & -\kappa^{*}_{~ U} \\
\kappa_{U} & -\kappa^{*}_{~ V} & \left( \dfrac{d}{dl} + \kappa_{I} \right) & \kappa^{*}_{~ Q} \\
\kappa_{V} &  \kappa^{*}_{~ U} & -\kappa^{*}_{~ Q} & \left( \dfrac{d}{dl} + \kappa_{I} \right) \end{array} \right \} 
\left \{ \begin{array}{c}
I_{\nu} \\[16pt]
Q_{\nu} \\[16pt]
U_{\nu} \\[16pt]
V_{\nu} \end{array} \right \} = 
\left \{ \begin{array}{c}
\eta_{\nu}^{~ I} \\[16pt]
\eta_{\nu}^{~ Q} \\[16pt]
\eta_{\nu}^{~ U} \\[16pt]
\eta_{\nu}^{~ V} \end{array} \right \} }
\end{equation}
\end{widetext}

\section{Polarized Radiative Transfer}

In order to compute both LP and CP at radio frequencies, we need a numerical algorithm to solve the full Stokes equations of polarized radiative transfer (summarized in \citealt{jones77,jones88}).  In particular, we solve the matrix presented in Equation \ref{eqn1} along various sight-lines (rays) passing through the TEMZ computational grid. Here, $I_{\nu}$, $Q_{\nu}$, $U_{\nu}$, and $V_{\nu}$ are the frequency-dependent Stokes parameters, while $(\eta_{\nu}^{~ I}, \eta_{\nu}^{~ Q}, \eta_{\nu}^{~ U}, \eta_{\nu}^{~ V})$ and $(\kappa_{I}, \kappa_{Q}, \kappa_{U}, \kappa_{V})$ represent, respectively, the emission and absorption coefficients for the corresponding Stokes parameters.  The terms $\kappa^{*}_{~ V}$ and $(\kappa^{*}_{~ Q}, \kappa^{*}_{~ U})$ account for the effects of Faraday rotation and conversion, respectively.  Finally, the path length through each TEMZ plasma cell is given by $l$.  There is an analytic solution to this matrix (presented in \citealt{jones77}) which we apply along individual rays passing through the TEMZ grid (depicted by a red line in Figure \ref{fig1}). We then monitor the polarized radiative transfer from cell to cell. 

We have integrated this polarized radiative transfer scheme into the ray-tracing code RADMC-3D.  This code builds up synthetic maps of the resultant polarized emission by casting thousands of rays through the TEMZ grid.  The exact analytic solutions to Equation \ref{eqn1} for $I_{\nu}$, $Q_{\nu}$, $U_{\nu}$, and $V_{\nu}$, as well as an expression for the optical depth $\tau$ of each homogenous plasma cell (i.e.~of uniform magnetic field strength), are given in \cite{jones77} and are summarized here in Appendix A.  RADMC-3D was designed to handle radiative transfer through non-relativistic  media.  Analytic expressions that account for the effects of relativistic aberration of the rays cast by RADMC-3D are outlined in \citep{lyutikov05} and have been incorporated into our algorithm (see \S \ref{Bonn}).  In \S \ref{Wardle} we present a test of our radiative transfer algorithm.

\begin{figure*}[!ht]
\begin{center}
\includegraphics[width=0.495\textwidth]{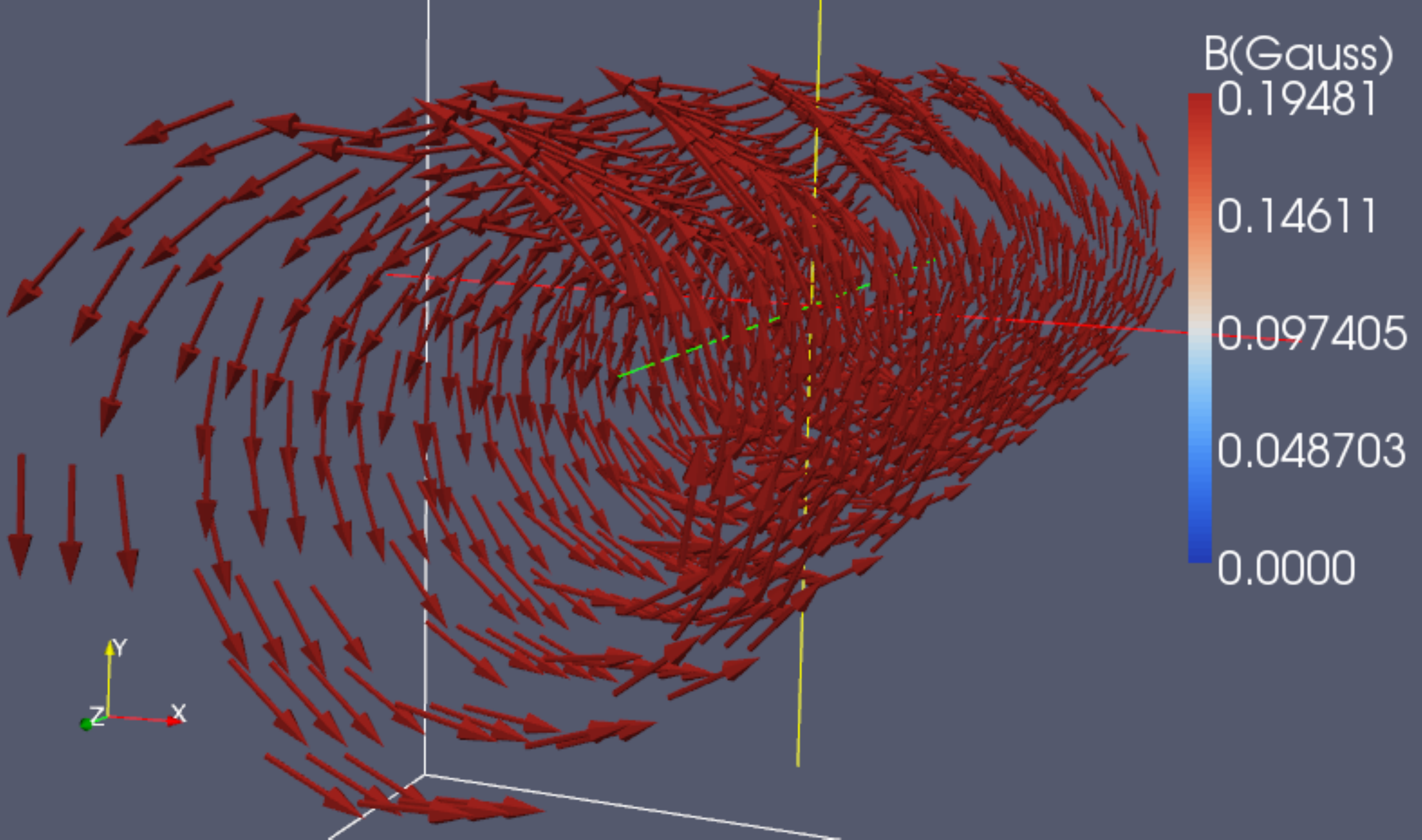}
\includegraphics[width=0.498\textwidth]{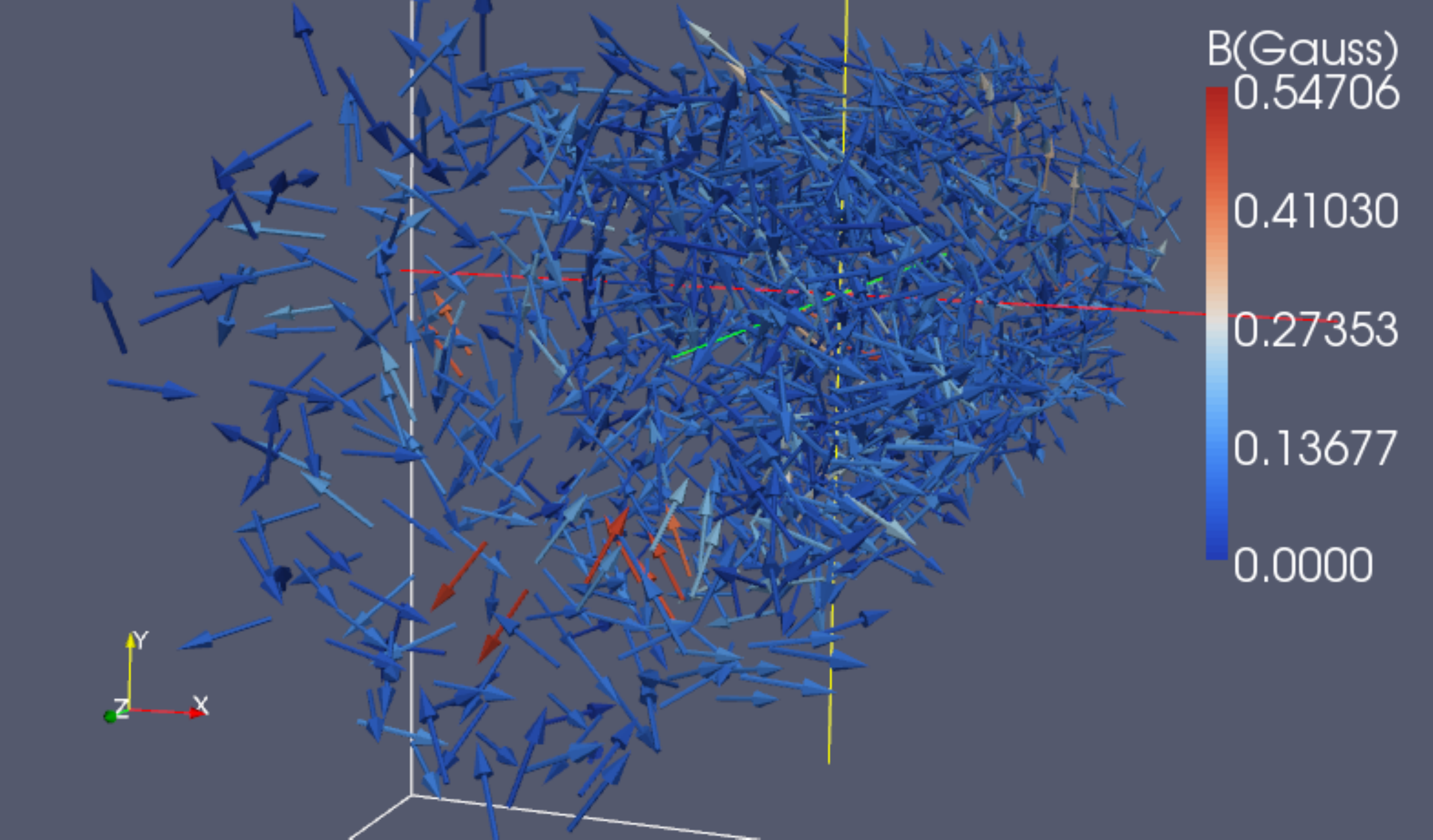}
\end{center}
\caption{(\textbf{Left panel}) A 3D visualization of the ordered helical magnetic field within the first TEMZ simulation. Each vector highlights the magnetic field strength within an~individual plasma cell (see color bar to the right for field strength in Gauss). (\textbf{Right panel}) A 3D visualization of the disordered magnetic field within the second TEMZ simulation.}
\label{fig2}
\end{figure*}

\section{Emission from an Ordered Magnetic Field}

\begin{figure*}
\centering
\includegraphics[width=0.55\linewidth]{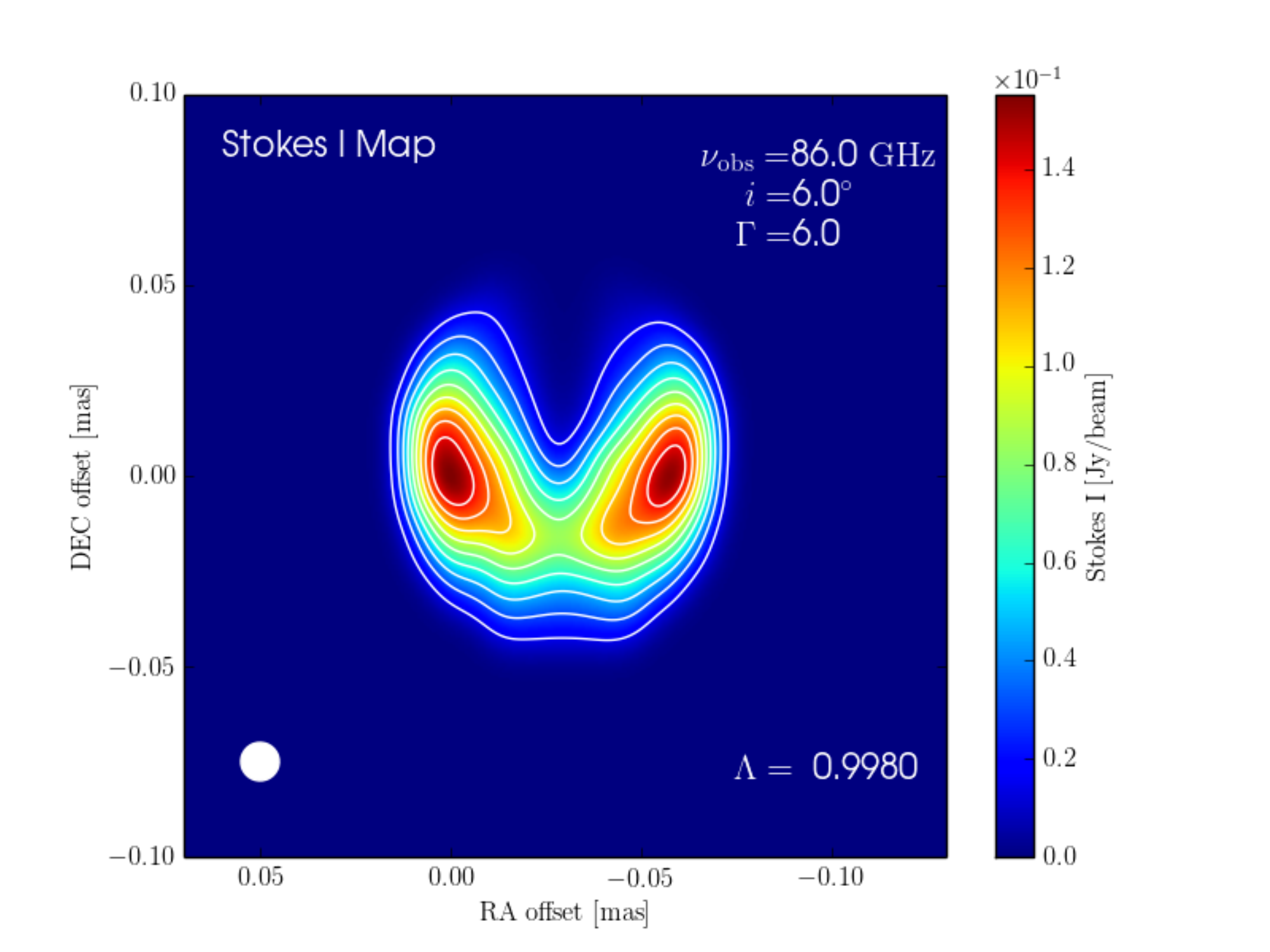}
\includegraphics[width=0.55\linewidth]{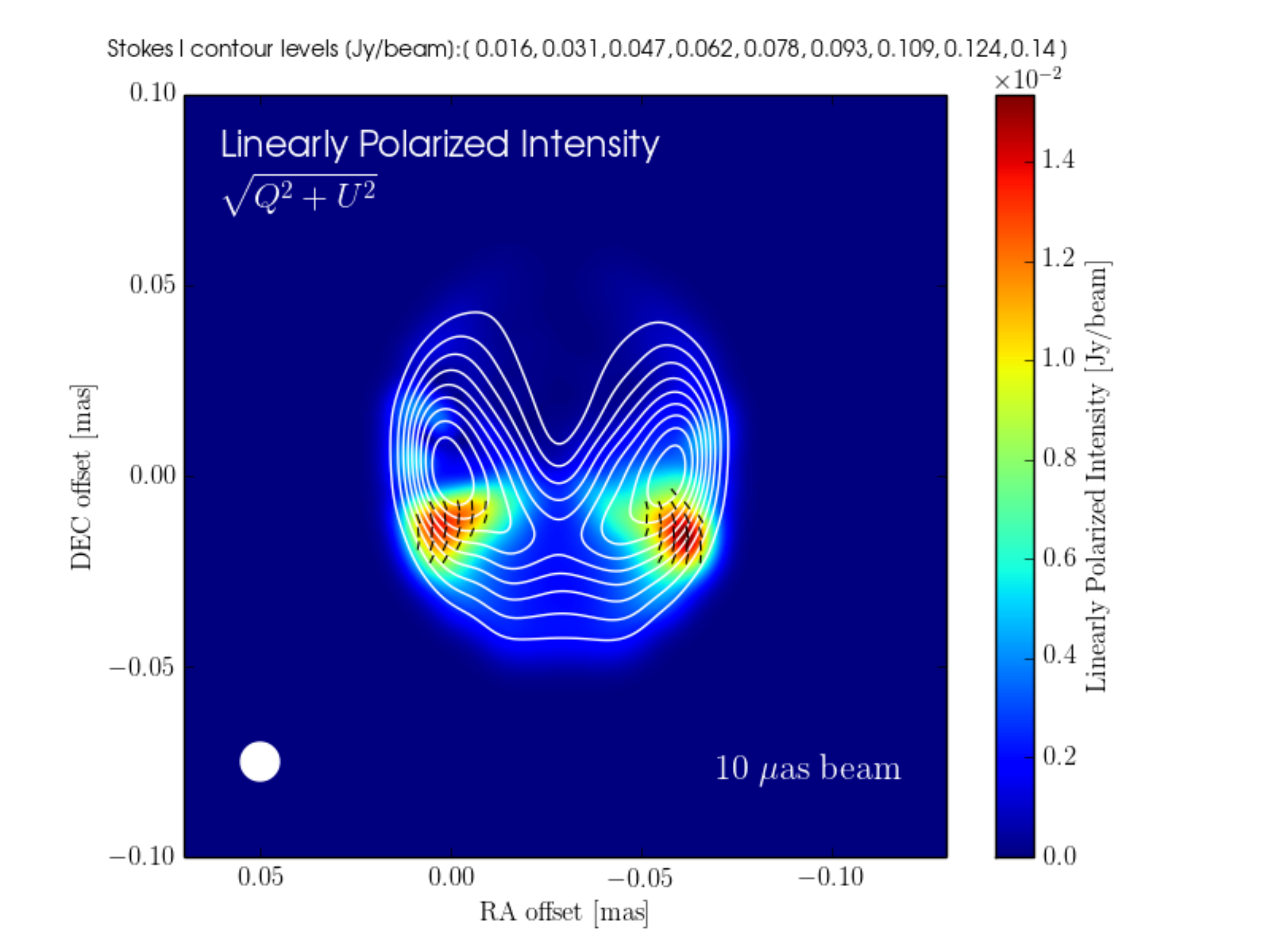}
\includegraphics[width=0.55\linewidth]{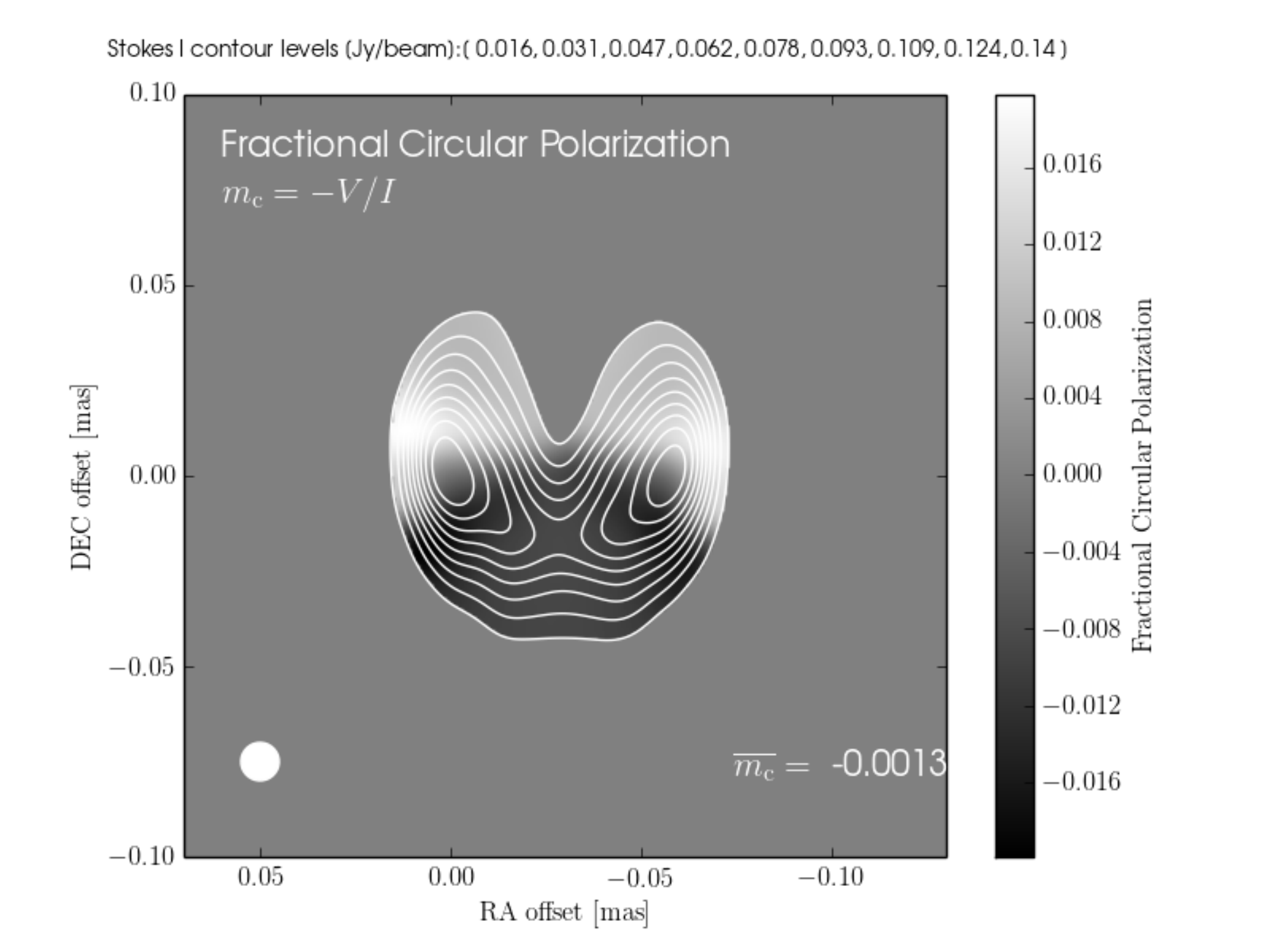}
\caption{\label{fig3}\textbf{(Upper panel)} - A rendering of the total intensity $I$ at an observing frequency of 86 GHz. The magnetic field has the ordered structure shown in the left panel of Figure \ref{fig2}. \textbf{(Middle panel)} - Rendering of linearly polarized intensity P of the ordered TEMZ grid.  Black line segments indicate the electric vector position angles (EVPAs) as projected onto the plane of the sky (I contours are overlaid in white). The effects of relativistic aberration (see \citealt{lyutikov05}) on the orientation of these EVPAs have been included in these calculations. \textbf{(Lower panel)} - A plot of the fractional CP in the observer's frame highlighting the different regions within the jet that produce positive and negative CP. An integrated value (see discussion in \S7) of CP is listed to the lower right. The above images have all been convolved with the circular Gaussian beam of FWHM $10 ~\mu\rm{as}$ shown in the lower left of each panel. $\Lambda$ (listed in the upper panel) pertains to the plasma composition of the jet and is discussed in \S10.}
\end{figure*}

Table \ref{tab1} lists the various adjustable parameters of the TEMZ model, which include the ratio of ordered magnetic field to the mean turbulent field. This ratio is very important to our present study because it determines the number of field reversals, shown by \cite{ruszkowski02} to play a dominant role in the generation of CP. As a trial calculation for our new polarized radiative transfer algorithm, we ray-traced through the highly ordered magnetic field depicted in the left panel of Figure \ref{fig2}.  For all of the calculations presented in this paper, the TEMZ grid is comprised of $13,440$ individual turbulent cells divided into eight concentric cylindrical shells creating the conical shock structure illustrated in Figure \ref{fig1}.  These cells of plasma were then mapped onto a uniform $120\times120\times120$ cartesian grid within which the ray-tracing calculation was carried out.  The RADMC-3D code casts $640,000$ individual rays through this cartesian grid forming $800\times800$ square-pixel maps of the polarized emission produced by the TEMZ model.  The RADMC-3D code makes use of a first-order cell-based scheme. The path length ($l$) through each turbulent TEMZ cell is computed and all plasma quantities are zone-centered. The resultant emission maps from this trial ray-tracing calculation are presented in Figure \ref{fig3}. We convolve the resultant images with a circular Gaussian beam with a full width at half maximum (FWHM) of $10 ~\mu\rm{as}$. This beam size is small enough to illustrate the main features of the model and also roughly corresponds to the angular resolution of the longest baselines of a global VLBI array (or one including an antenna in space) at short radio wavelengths for comparison with potential future observations.

\begin{deluxetable}{llc}
\tablecolumns{3}
\tablewidth{1.0\columnwidth}
\tabletypesize{\normalsize}
\tablecaption{TEMZ Model Parameters and Values Used in this Study \label{tab1}}
\startdata
\tableline
\noalign{\smallskip}
\tableline
\noalign{\smallskip}
Symbol & Description &Value \\
\tableline
\noalign{\smallskip}
$Z$ & Redshift & 0.069 \\
\noalign{\smallskip}
$\alpha$ & Optically thin spectral index & 0.65 \\
\noalign{\smallskip}
$B$ & Mean magnetic field (G) & 0.03 \\
\noalign{\smallskip}
$-b$ & Power-law slope of power spectrum & 1.7 \\
\noalign{\smallskip}
$f_{B}$ & Ratio of energy densities $u_{e}/u_{B}$ & 0.1 \\
\noalign{\smallskip}
$R_{\rm{cell}}$ & Radius of TEMZ cells (pc) & 0.004 \\
\noalign{\smallskip}
$\gamma_{\rm{max}}$ & Maximum electron energy & 10000  \\
\noalign{\smallskip}
$\gamma_{\rm{min}}$ & Minimum electron energy & 10 \\
\noalign{\smallskip}
$\beta_{\mathbf{u}}$ & Bulk laminar velocity of unshocked plasma & 0.986 \\
\noalign{\smallskip}
$\beta_{\mathbf{t}}$ & Turbulent velocity of unshocked plasma & 0.0 \\
\noalign{\smallskip}
$\zeta$ & Angle between conical shock and jet axis & $10^{\circ}$ \\
\noalign{\smallskip}
$\theta_{\rm{obs}}$ & Angle between jet axis and line of sight & $6^{\circ}$ \\
\noalign{\smallskip}
$\phi$ & Opening semi-angle of jet & $6.8^{\circ}$ \\
\noalign{\smallskip}
$z_{\rm{MD}}$ & Distance of Mach disk from BH (pc) & 1.0 \\
\noalign{\smallskip}
$f_{\rm{field}}$ & Ratio of ordered helical field to total field & 0.1-1.0 \\
\noalign{\smallskip}
$\psi$ & Pitch angle of the helical field & $45^{\circ}$  \\
\noalign{\smallskip}
$n_{\rm{step}}$ & Number of time steps & 5000 \\
\noalign{\smallskip}
$n_{\rm{rad}}$ & Number of cells in jet cross-section & 169
\enddata
\end{deluxetable} 

The jet was given a bulk Lorentz factor of $\Gamma = 6$ and an angle of inclination to our line of sight of $6^{\circ}$.  Each computational cell in our model has a characteristic length scale of $0.004$ pc.  An underlying electron power-law energy distribution, $n_{e}(\gamma) \propto \gamma^{-s}$, was assigned to each cell over the energy range $\gamma_{\rm min}$ to $\gamma_{\rm max}$, with $s=2.3$ applied throughout the whole grid. The low-energy cutoff ($\gamma_{\rm{min}}$) of the electron energy distribution has a major influence on the emission we compute from our models. We adopt a value of $\gamma_{min} = 10$ which corresponds to the high rotation limit (i.e.~nearly circular characteristic waves, discussed further in \S5 \& \S11). After crossing the shock, each cell's values of $\gamma_{\rm min}$ and $\gamma_{\rm max}$ are updated as the electrons undergo radiative cooling. This cooling is computed analytically and takes account of the local magnetic field strength and the density of seed photons (Equation \ref{eqn0}). As the plasma propagates down the cell structure, the history of the losses is stored, so that the losses during a given time step are incrementally added to the previous losses in a self-consistent fashion (see \citealt{marscher14} - \S2.1 for a more in-depth discussion). We also point out that the ratio of energy densities ($u_{e}/u_{B}$) listed in Table \ref{tab1} is applied upstream of the shock and evolves down the length of the jet as well. Table \ref{tab1} lists the values of the other TEMZ model parameters used in our calculations.  Figure \ref{fig3} (lower panel) shows a map of the fractional circular polarization ($m_{\rm c} \equiv -V/I$) emerging from the TEMZ grid.  Unlike observations that typically show only one sign of CP in a given blazar, our synthetic map contains both positive and negative CP that appear simultaneously.  The levels of circular polarization in this map reach $\sim 1.6\%$.  The highest levels of circular polarization observed to emanate from an extragalactic jet are $\sim 2$ - $4\%$ in the radio galaxy 3C~84 (\citealt{homan04}).  While these initial results are promising, the magnetic field used here is highly idealized.

\section{Faraday Conversion in a Disordered Magnetic Field}

\begin{figure}
  \setlength{\abovecaptionskip}{-6pt}
  \begin{center}
  \hspace*{-0.65cm}
    \scalebox{0.94}{\includegraphics[width=1.15\columnwidth,clip]{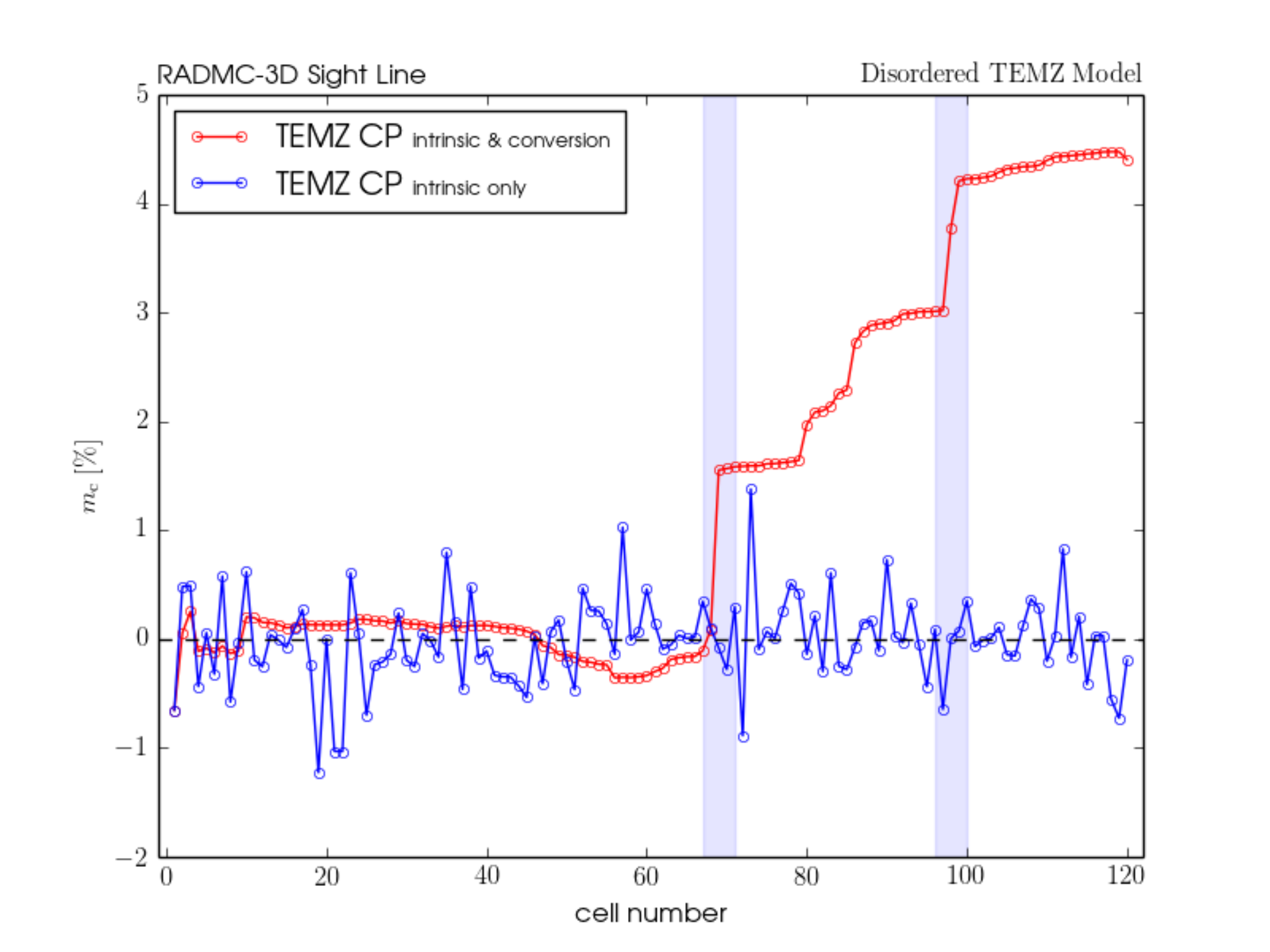}}
   \hspace*{-0.65cm}
    \scalebox{0.94}{\includegraphics[width=1.15\columnwidth,clip]{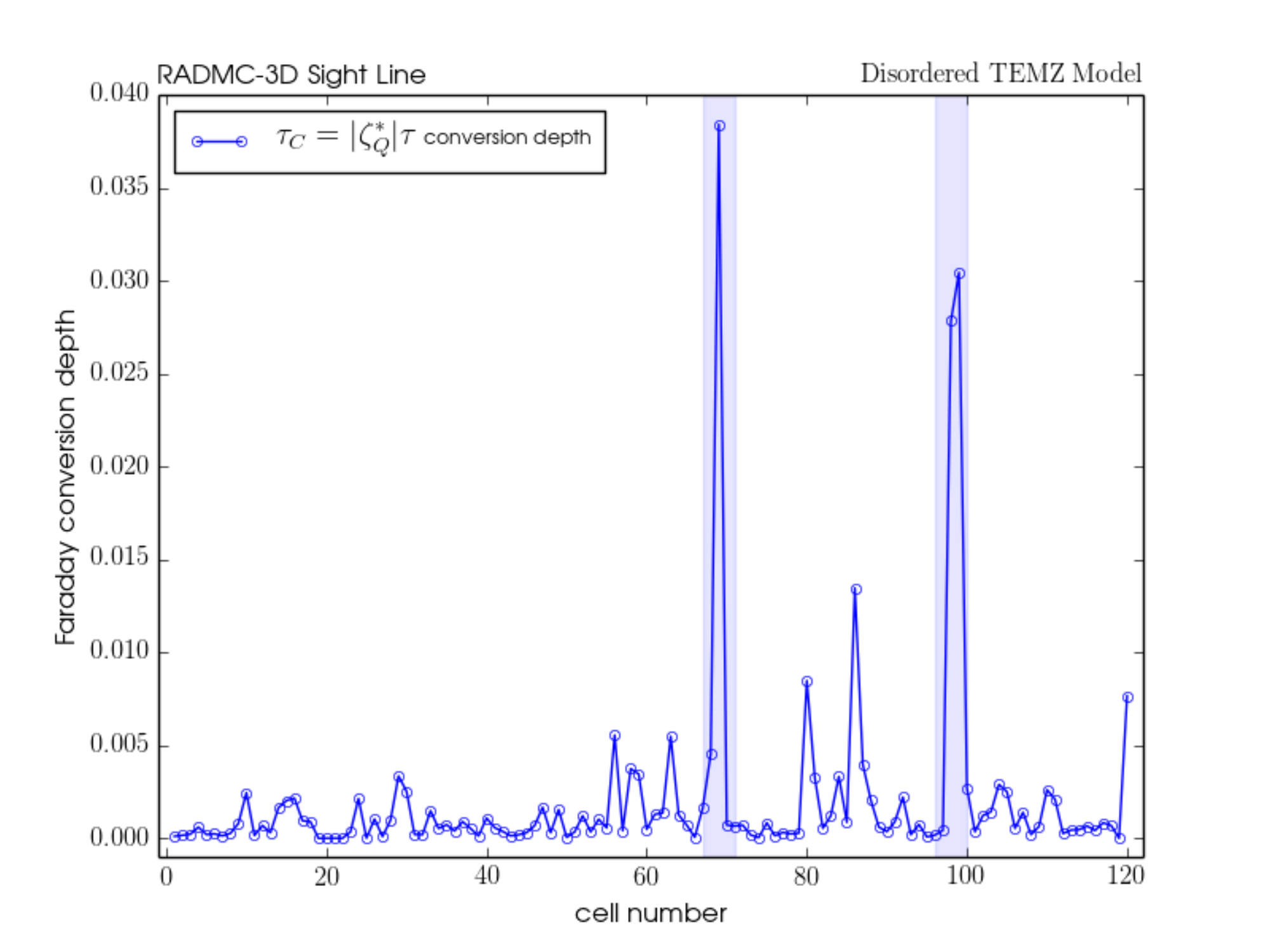}}
  \end{center}
  \caption{\label{fig4} \textbf{(Upper panel)} - A plot of fractional circular polarization $m_{\rm c}$ along a random sight-line through the disordered TEMZ grid.  The blue circles correspond to CP intrinsic to each TEMZ cell, whereas the red circles highlight the effect that Faraday conversion has on the observed level of CP along the ray.  The radiative transfer starts at the back of the jet (cell 0) and progresses to the front of the jet (cell 120), resulting in the creation of one pixel in our image maps.  \textbf{(Lower panel)} - A plot of the Faraday conversion depth $\tau_{C}$ (blue circles) along the same sight-line presented in the upper panel.  The cell numbers are identical.}
\end{figure}

After performing the initial ray-tracing calculation presented in \S4, we applied our scheme to the case of a turbulent magnetic field (illustrated in the right panel of Figure \ref{fig2}). One would expect that the increased disorder in the field would increase the birefringence of the jet plasma, leading to increased Faraday conversion, as found by \cite{ruszkowski02}. It might also cause depolarization of the emission due to the lack of a systematic vector-ordered magnetic field component.  The interplay between these two effects should govern the level of CP produced in the turbulent TEMZ grid.  In order to explore these effects in a quantifiable manner, we have tracked several emission parameters pertaining to CP production along a random sight-line through this disordered TEMZ model.  As a first test, we recorded the fractional CP ($m_{\rm c}$) from cell to cell along a given sight-line (see upper panel of Figure \ref{fig4}).  In one test case (shown in blue in the upper panel of Figure \ref{fig4}), we plot only the fractional CP intrinsic to each cell.  In another test case (shown in red in the upper panel of Figure \ref{fig4}), we also allow Faraday conversion to modify the emission from cell to cell along our sight-line.  One can clearly see that as our ray passes through the turbulent plasma, Faraday conversion increases the observed level of $m_{\rm c}$ beyond what would be intrinsic to each cell.  We also call attention to the high levels of $m_{\rm c}$ that are produced within the disordered TEMZ model ($\sim 4.5 \%$ for this particular sight-line).  These ``extreme'' CP pixels get ``washed'' out when we convolve our resolved emission maps with a circular Gaussian beam resulting in integrated levels of CP $m_{\rm c} \lesssim 1\%$ (see \S6 \& \S8).

As summarized in \cite{wardle03} and \cite{ruszkowski02}, the production of CP within blazars can result from Faraday conversion in a turbulent jet plasma. As discussed in \S2 and \S3, we model this turbulent plasma using thousands of individual homogeneous plasma cells. We point out, however, that \cite{hodge82} found that variations in the magnetic field of an inhomogeneous plasma can lead to characteristic wave coupling which in turn produces sizable amounts of CP. \cite{bjornsson90} further showed that the coupling of characteristic waves can occur over smooth variations of the plasma properties. This implies that the plasma properties, which in a homogeneous plasma cell create CP, in an inhomogeneous source would also produce coupling between the characteristic waves. The CP produced through this coupling is predicted to be of the same order as that from Faraday conversion (\citealt{bjornsson90}). Therefore, our modeling a turbulent jet as a piecewise homogeneous structure artificially excludes an important source of CP. To include these complex plasma effects, however, is beyond the scope of this paper.  

Faraday conversion occurs over two stages. Initially, linearly polarized emission originates from within the jet via synchrotron emission from relativistic electrons entrained in the jet's magnetic fields.  If we choose a suitable reference axis, this process will generate Stokes Q.  Internal Faraday rotation can then convert Q into U as the synchrotron radiation propagates through the jet. In our simulation the cells in the TEMZ model act as internal Faraday screens (see \S9).  Faraday conversion, a process uniquely intrinsic to relativistic plasma, then transforms U into V.  Therefore, in our model the production of CP within a blazar jet is dependent upon both the Faraday rotation depth ($\tau_{F}$) and the Faraday conversion depth ($\tau_{C}$) through the intervening jet plasma along each line of sight. These Faraday depths are themselves dependent upon the optical depth ($\tau$) of the jet plasma:
\begin{equation}\label{eqn2}
\tau_{F} = |\zeta_{V}^{*}| ~ \tau
\end{equation}
and
\begin{equation}\label{eqn3}
\tau_{C} = |\zeta_{Q}^{*}| ~ \tau ~ ,
\end{equation}
where $\zeta_{V}^{*}$, $\zeta_{Q}^{*}$, and $\tau$ are given in Appendix A; their values are specific to the synchrotron process (see Equations \ref{eqnA2}, \ref{eqnA11}, and \ref{eqnA12}, respectively). In a homogeneous cell of plasma (i.e.~of uniform magnetic field) the observed levels of fractional CP from the conversion process are proportional to the ratio of the Faraday depths: $m_{\rm c} \propto ~ \tau_{C}/\tau_{F}$ in the high rotation ($\tau_{F} \gg 1$) limit (\citealt{jones77}) and the product of the Faraday depths: $m_{\rm c} \propto \tau_{F} ~ \tau_{C}$ in the low rotation ($\tau_{F} \ll 1$) limit (\citealt{wardle03}, \citealt{homan09}, \& \citealt{osullivan13}). We explore the spectral behavior of $m_{\rm c}$ in these two limiting cases in \S11. As a second test of CP production in the TEMZ model, in the lower panel of Figure \ref{fig4} we plot the Faraday conversion depth from cell to cell along the same sight-line shown in the upper panel of Figure \ref{fig4}.  It is clear from comparison of the upper and lower panels of Figure \ref{fig4} (highlighted in blue) that the two spikes in the conversion depth of the plasma along this particular sight-line result in corresponding increases in $m_{c}$. An additional test of our CP algorithm is presented in \S \ref{Wardle}.

\section{Emission from a Disordered Magnetic Field}

\begin{figure*}
\centering
\includegraphics[width=0.55\linewidth]{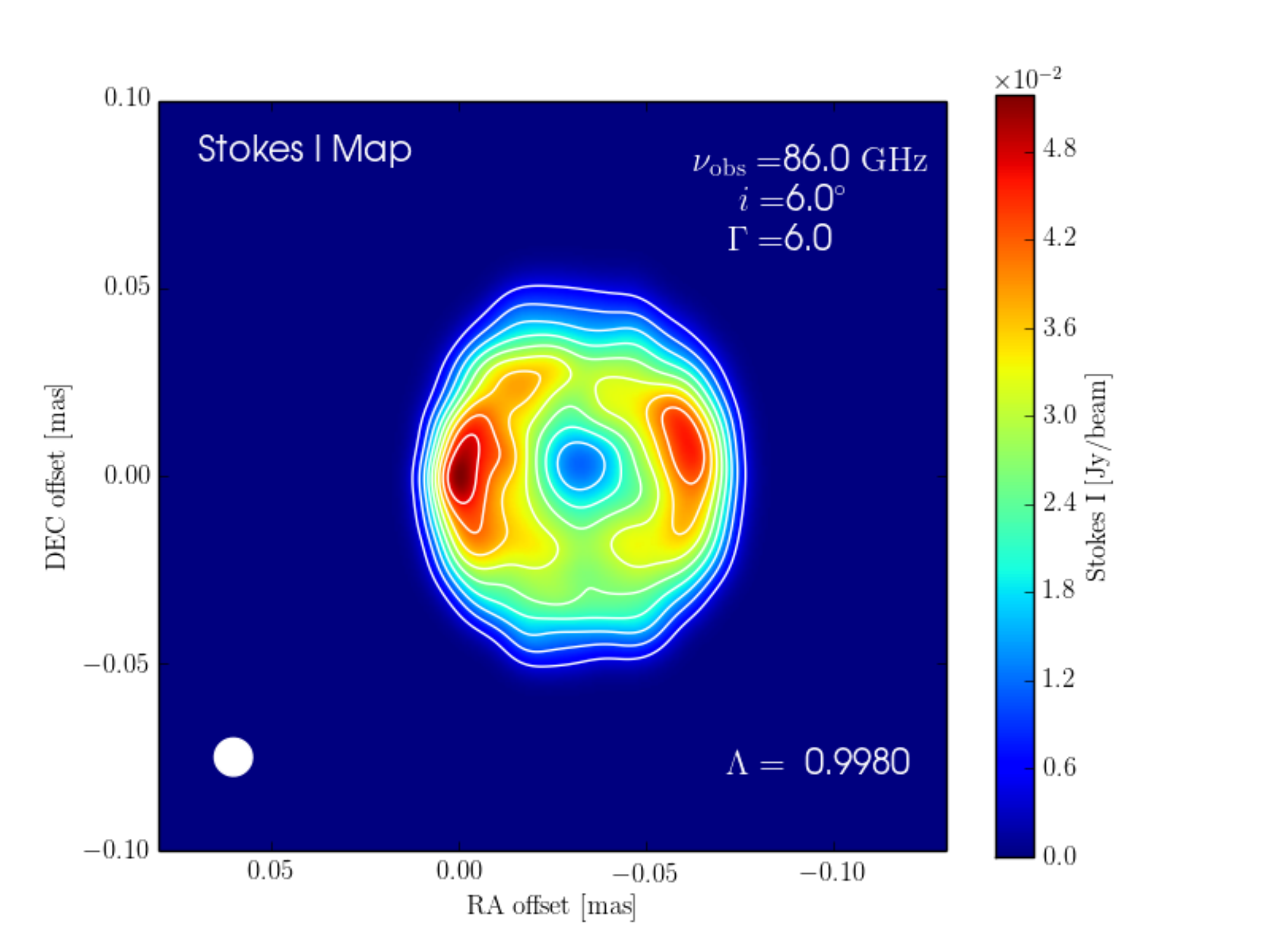}
\includegraphics[width=0.55\linewidth]{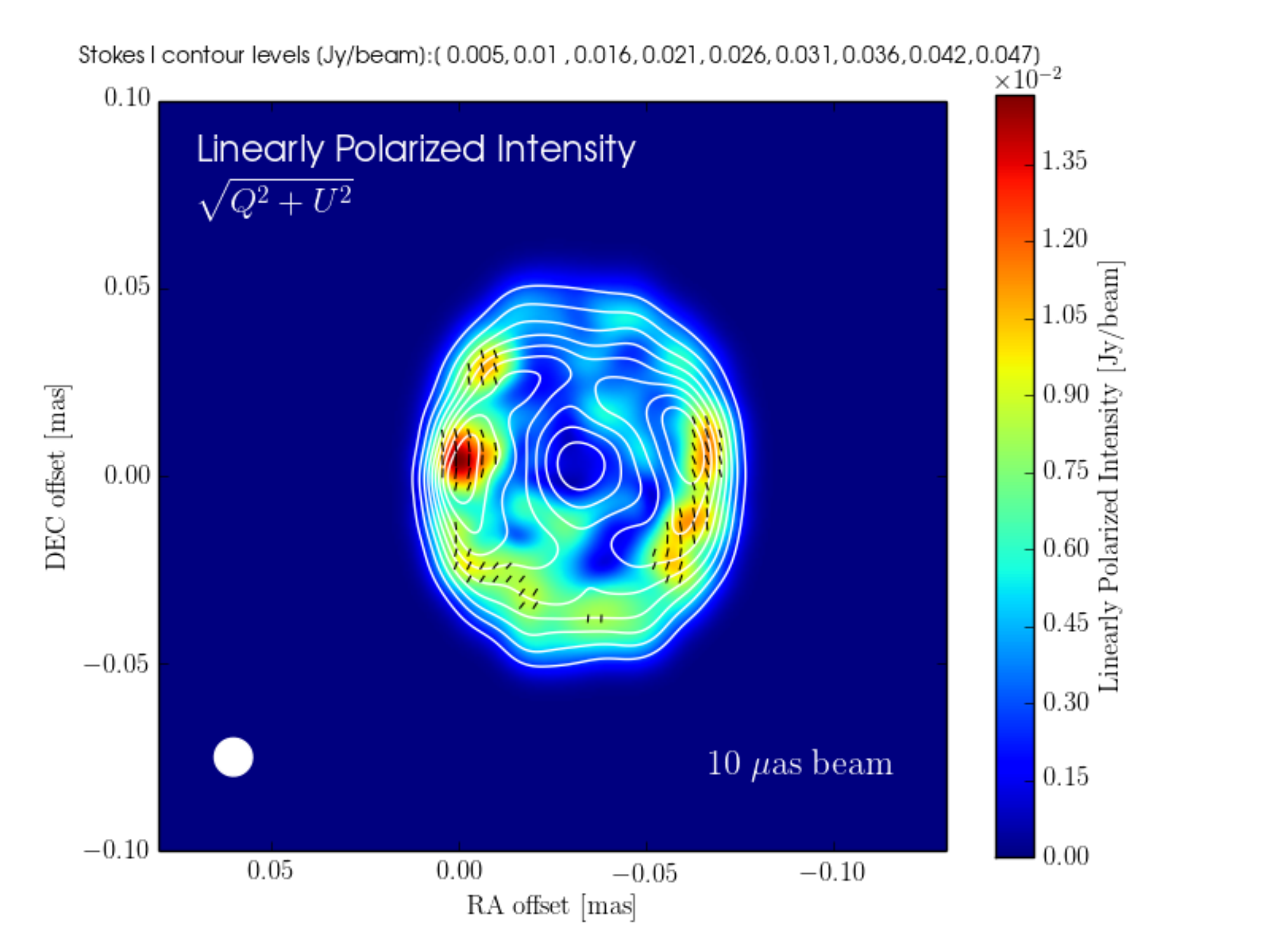}
\includegraphics[width=0.55\linewidth]{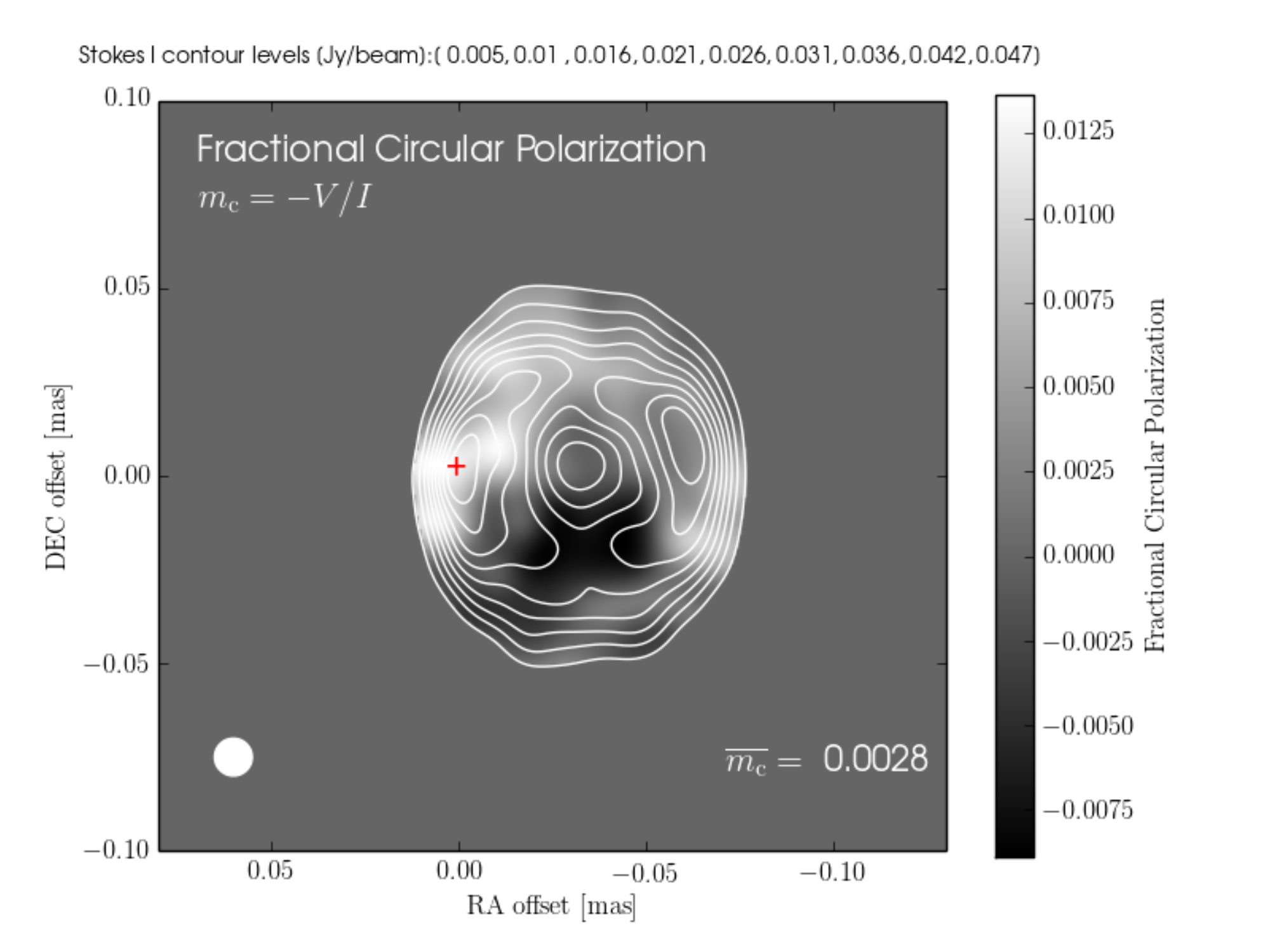}
\caption{\label{fig5} \textbf{(Upper panel)} - A rendering of the total intensity $I$ at an observing frequency of 86 GHz. The magnetic field has the disordered structure shown in the right panel of Figure \ref{fig2}. \textbf{(Middle panel)} - Rendering of linearly polarized intensity P of the disordered TEMZ grid.  Black line segments indicate the electric vector position angles (EVPAs) as projected onto the plane of the sky (I contours are overlaid in white). The effects of relativistic aberration (see \citealt{lyutikov05}) on the orientation of these EVPAs have been included in these calculations. \textbf{(Lower panel)} - A plot of the fractional CP in the observer's frame highlighting the different regions within the jet that produce positive and negative CP. The red cross demarcates the sight-line examined in Figure \ref{fig4}. An integrated value (see discussion in \S7) of CP is listed to the lower right. The above images have all been convolved with the circular Gaussian beam of FWHM $10 ~\mu\rm{as}$ shown in the lower left of each panel. $\Lambda$ (listed in the upper panel) pertains to the plasma composition of the jet and is discussed in \S10.}
\end{figure*}

We proceed to explore the effect that this more turbulent jet environment has on the CP emission maps produced by RADMC-3D. All of the model parameters listed in Table \ref{tab1} remain the same as for the ordered field computations, except here $f_{\rm field}$, the ratio of the ordered helical magnetic field to the total magnetic field within each TEMZ cell, is set to 0.1. As discussed in \S5, the expectation is that the increased disorder in the turbulent TEMZ magnetic field will lead to increased levels of CP. The images from this second ray-trace (shown in Figure \ref{fig5}) illustrate the more disordered emission (as compared to Figure \ref{fig3}) produced by this calculation. We again convolve the resultant images with a circular Gaussian beam of FWHM $10 ~\mu\rm{as}$. We also convolve the images (shown in Appendix B) with an elliptical Gaussian beam of FWHM~$0.14 \times 0.05$ mas which mimics the resolution currently attainable with the Global Millimeter VLBI array (GMVA) at 86 GHz.

The integrated level of circular polarization in the turbulent TEMZ model ($\overline{m_{\rm c}}$, listed in the lower panel of Figure \ref{fig5}, and discussed in detail in \S7) reaches $\sim$0.3\%, which is still in rough agreement with observations of circular polarization in blazars (see \citealt{homan04}).  However, as discussed in \S5, we find very localized levels of fractional circular polarization that can reach as high as $\sim$4\% in a disordered (upstream of the shock) TEMZ field. In all instances the fractional LP ($m_{\rm l} \equiv \sqrt{ Q^{2} + U^{2} }/I$) is much larger than the absolute value of the fractional CP ($|m_{\rm c}| = |V|/I$); however, $m_{\rm l}$ always remains below $\sim 70\%$ (the theoretical maximum for synchrotron emission with $\alpha=0.65$; \citealt{pacholczyk70}). The images presented in Figure \ref{fig5} are obtained at an observing frequency of $\nu_{\rm obs} = 86 ~ \rm{GHz}$ and correspond to optically thin synchrotron emission. The I and P maps are positive definite. The $m_{\rm c}$ map exhibits a range of positive and negative values. A study of the temporal and spectral variability of these synthetic images is presented in \S8 and \S11.

The middle panels of Figures \ref{fig3} and \ref{fig5} present maps of linearly polarized intensity ($P \equiv \sqrt{ Q^{2} + U^{2} }$).  The orientation of the LP as projected onto the plane of sky is demarcated by electric vector position angles (EVPAs) $\chi$.  These EVPAs are shown as black line segments in the P maps.  The value of $\chi$ within a given pixel is computed based on the local values of Q and U within that pixel: $\chi = \frac{1}{2} ~ \rm{arctan}( ~ U/Q ~ )$.  As a further test of our algorithm, we tune our observing frequency until the emission emanating from the TEMZ grid reaches the optically thick limit.  As $\tau \rightarrow 1$ we observe the polarization vectors to systematically rotate by $\sim 90^{\circ}$, as predicted by the theory of synchrotron self-absorption (\citealt{pacholczyk70}; see the variation of the integrated values of $\chi$ with $\nu$ in Figure \ref{fig9}).  Images obtained of the polarized intensity in the co-moving frame of the plasma clearly show ordering of the initially turbulent magnetic field downstream of the shock. 

For all of the calculations presented in this paper, the angle of the jet to our line of sight ($\theta_{\rm obs}$) is set to $6^{\circ}$.  All of our model parameters are similar to those used by \cite{wehrle16} to model BL Lacertae.  In the future, we plan to carry out a more detailed study of the effect that varying degrees of jet inclination can have on the observed levels of CP.  This test will be crucial in determining whether the sign of $m_{\rm c}$ is sensitive to the large scale orientation of the jet's magnetic field with respect to our line of sight (see \citealt{ensslin03,ruszkowski02}). 

\section{Integrated Levels of Fractional Polarization}

At the resolution of VLBI polarized intensity images published thus far, the structures displayed in Figures \ref{fig3} and \ref{fig5} would only be partially resolved or even unresolved.  In order to make comparisons between the levels of CP present in our simulations and existing measurements, we therefore compute integrated levels of fractional CP at each time step/epoch of the simulation.  This is accomplished by first defining a limiting contour in our I maps (which we arbitrarily set to 10\% of the peak I value).  We then compute $m_{\rm c} = - V/I$ in each pixel within this limiting contour, creating the maps of fractional circular polarization presented in the lower panels of Figures \ref{fig3} and \ref{fig5}.  To acquire integrated levels of fractional CP, we then compute averages of the Stokes I and V parameters within this limiting contour: $\overline{I}$ and $\overline{V}$.  Based on these averages, we then compute an integrated level of fractional CP for each time step of the simulation: 
\begin{equation}\label{eqn4}
\hspace*{-1.2cm}\overline{ m_{\rm c} } = - \overline{V}/\overline{I} ~ .
\end{equation} 
As discussed in \S6, the levels of fractional CP present in Figures \ref{fig3} and \ref{fig5} are in rough agreement with observations of CP.  In \S8, we carry out a temporal analysis in which we monitor these integrated levels of fractional CP over the course of the entire TEMZ simulation.  The presence of both negative and positive levels of CP (simultaneously) within our maps is a distinct feature that has not been commonly seen in observations of blazar jets (with the exception of 3C 84 - see \citealt{homan04}), but this may be the result of the cores only being partially resolved by VLBI observations.  It will be of interest to compare our synthetic images with the results of upcoming observational campaigns that will image several blazars on $\mu\rm{as}$ scales with various VLBI arrays, including RadioAstron plus ground based antennas, the Global Millimeter VLBI Array (GMVA), and the Event Horizon Telescope (EHT).  The sensitivity and resolution of these combined arrays should be sufficient to probe the location(s) in the jet where Faraday conversion occurs.

\begin{figure}
  \setlength{\abovecaptionskip}{-6pt}
  \begin{center}
  \hspace*{-0.65cm}
    \scalebox{0.96}{\includegraphics[width=1.15\columnwidth,clip]{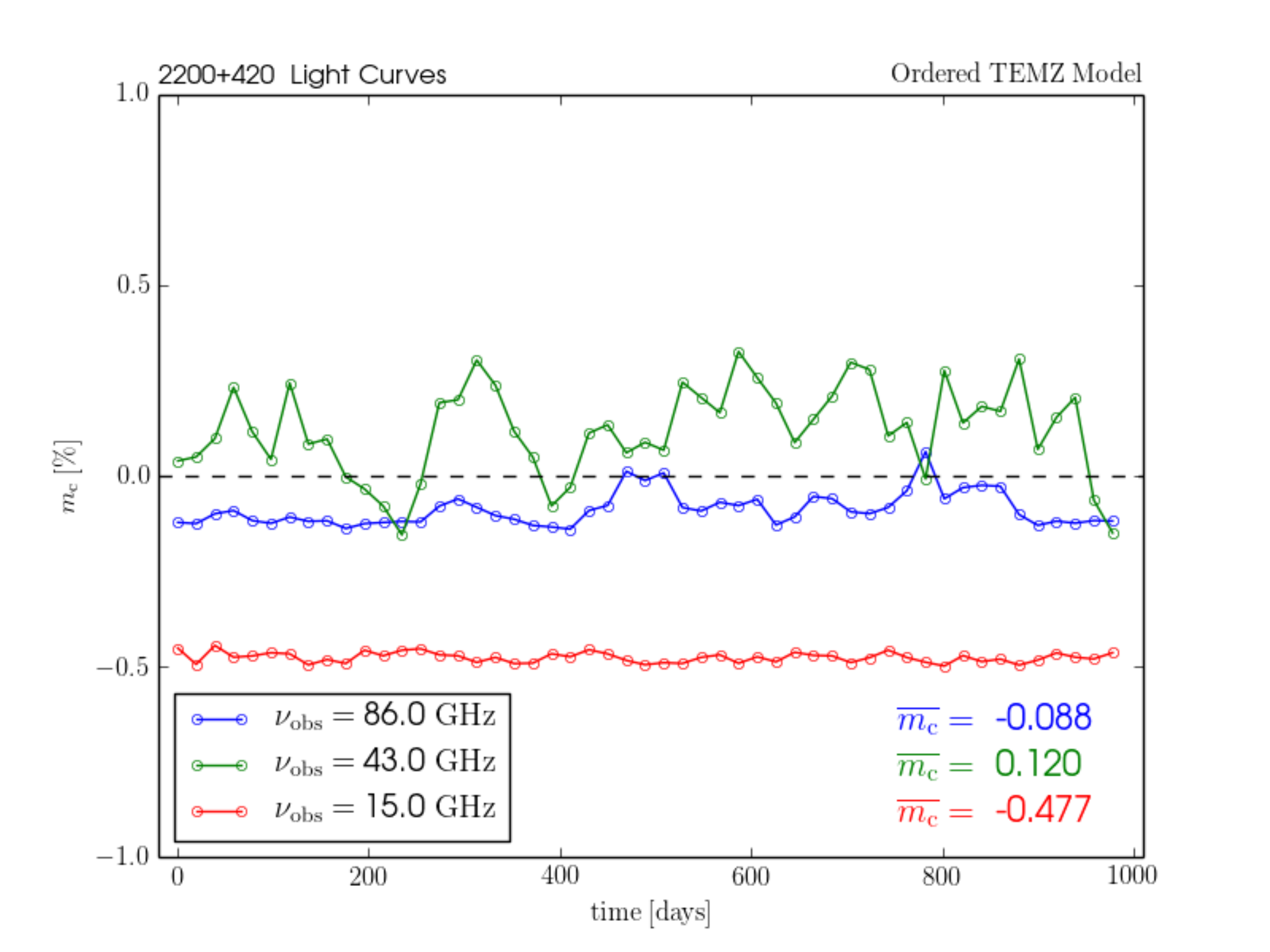}}
   \hspace*{-0.65cm}
    \scalebox{0.96}{\includegraphics[width=1.15\columnwidth,clip]{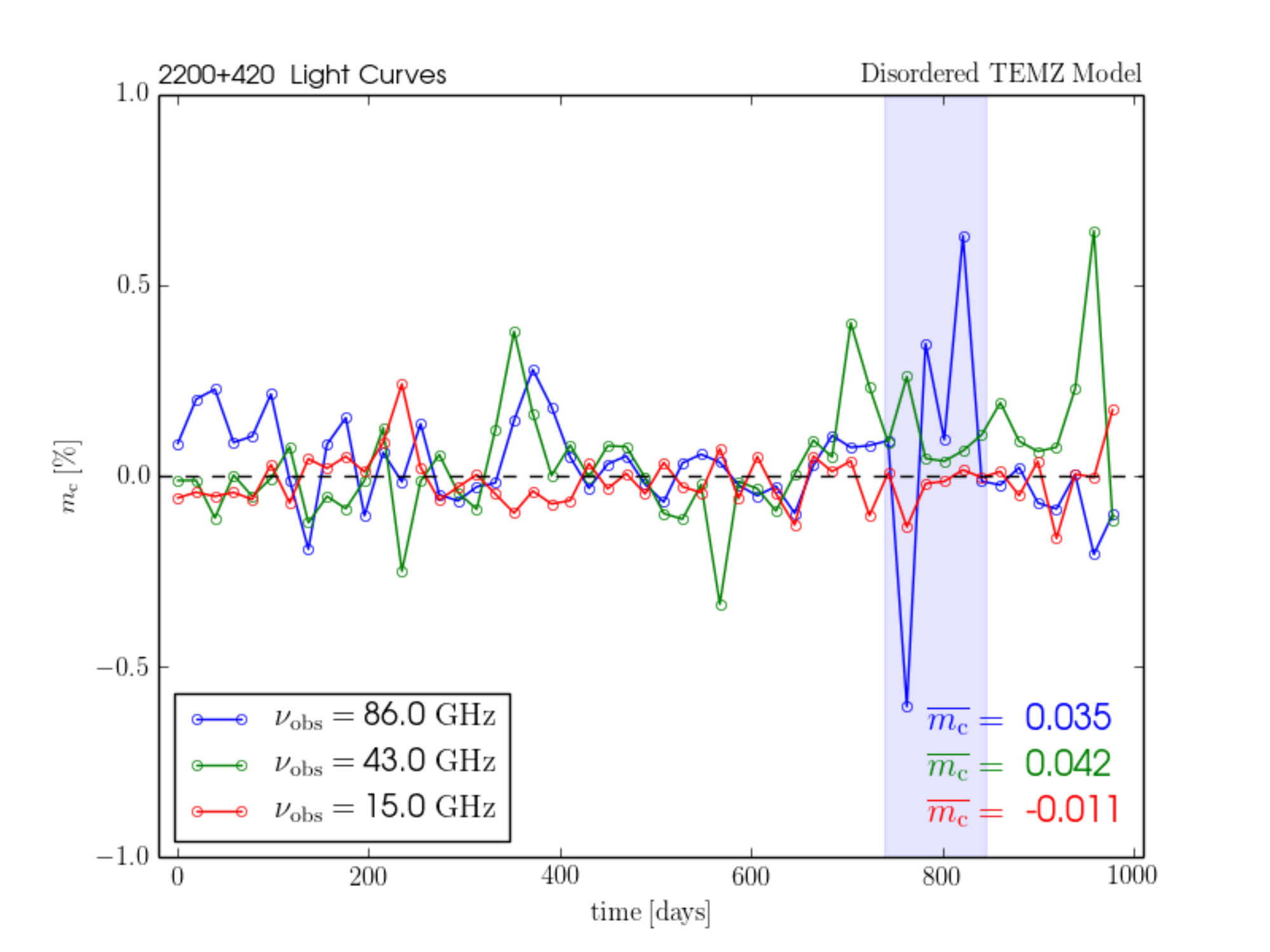}}
  \end{center}
  \caption{ \label{fig6} \textbf{(Upper panel)} - Variations of $m_{\rm c}$ for the ordered field case (left panel of Figure \ref{fig2}).  The simulation is run for 5000 time steps, with each time step corresponding to $\sim 0.2$ days in the observer's frame.  The red, green, and blue curves correspond to observations at $\nu_{\rm obs} =$ 15, 43, and 86 GHz, respectively.  The average values of $m_{\rm c}$ for each frequency over the entire duration of the simulation are listed in the lower right. \textbf{(Lower panel)} - Corresponding variability in $m_{\rm c}$ for the disordered field case (right panel of Figure \ref{fig2}). The blue bar highlights a CP sign reversal at 86 GHz.}
\end{figure}

\section{Multi-Epoch Observations}

There have been several multi-frequency studies of the temporal variability of CP observed in blazar jets.  In particular, the University of Michigan Radio Astronomy Observatory (UMRAO) has carried out several dedicated programs with single 26 m dish observations at frequencies of 4.8, 8.0, and 14.5 GHz (see, e.g., \citealt{aller11}; \citealt{aller03}).  These campaigns have detected CP at faint but statistically significant levels in a number of blazar jets (e.g., 3C 273, 3C 279, 3C 84, and 3C 345).  The POLAMI program (see \citealt{thum18}) is also actively conducting CP monitoring of blazars at mm-wavelengths. A marked feature in the temporal variability of CP in these sources is the occurrence of sign reversals in the handedness of the observed fractional CP.  These sign changes occur over a range of timescales (month to years).  It has been suggested that these sudden reversals in the sign of $m_{\rm c}$  indicate the existence of turbulent magnetic fields within the radio cores of these objects (see \citealt{aller03}).  \cite{ensslin03} point out, however, that there are long time spans (decades in some cases) over which the sign of $m_{\rm c}$ can remain constant (see also \citealt{myserlis18}).  \citeauthor{ensslin03} suggested that in these objects Faraday conversion occurs as the radio waves propagate through a rotationally twisted rather than turbulent magnetic field.  Since the twist could result from rotation of the field lines at the base of the jet (e.g., if the footprint of the field is in plasma orbiting the black hole), CP sign reversals would be quite rare.  These two physical scenarios highlight the potential of using observations of temporal variability in CP as a probe of the underlying physics of relativistic jets.

In order to investigate these two scenarios, we have carried out a time series analysis of the integrated levels of fractional CP produced by the TEMZ model using two different magnetic field grids: (i) the ordered helical magnetic field geometry shown in the left panel of Figure \ref{fig2}, and (ii) the turbulent magnetic field geometry shown in the right panel of Figure \ref{fig2}.  We run each TEMZ model for 5000 time steps.  Each time step (as outlined in \citealt{marscher14}) is determined by how long it takes one plasma cell to advect down the jet by one cell length (see Figure \ref{fig1}).  With a bulk Lorentz factor $\Gamma = 6$ and viewing angle of $6^\circ$ (Table 1), this corresponds to $\sim0.2$ days in the observer's frame.  All images in our time series analysis have been convolved with a circular Gaussian beam with a FWHM of $10 ~\mu\rm{as}$. Finally, as discussed in \S7, we compute integrated levels of fractional circular polarization $\overline{m_{\rm c}}$ based on averages of I and V computed within the limiting contours of each map.  These integrated values are then recorded from one time step to the next, creating synthetic multi-epoch observations.  We perform this analysis for both grids at frequencies of $\nu_{\rm{obs}} =$ 15, 43, and 86 GHz.  The results are shown in Figure \ref{fig6}, in which the upper panel illustrates the temporal variability in $\overline{m_{\rm c}}$ for the ordered field case, whereas the lower panel highlights the corresponding variability in $\overline{m_{\rm c}}$ for a disordered field.  The average value of $\overline{m_{\rm c}}$ for each TEMZ simulation is listed (for each frequency) in the lower right of each panel.  The sign of $m_{\rm c}$ over multiple epochs in the ordered field case, is in keeping with the predictions of \cite{ruszkowski02} and \cite{ensslin03}.  In contrast, when the field is disordered, we indeed see successive reversals in the sign of $\overline{m_{\rm c}}$ (one such reversal at 86 GHz, occurring over roughly 100 days, is highlighted in blue in the lower panel of Figure \ref{fig6}).  At $\nu_{\rm{obs}} =$ 15 GHz the value of $\overline{m_{\rm c}}$ is predominantly negative for both runs, whereas at $\nu_{\rm{obs}} =$ 43 GHz the value of $\overline{m_{\rm c}}$ tends to be positive.  In \S11 we carry out a more detailed analysis of the spectral behavior of $m_{\rm c}(\nu)$.  The temporal variability of $m_{\rm c}$ for the disordered field supports the interpretation proposed by \cite{aller11} that reversals in the sign of CP are an observational signature of turbulence within the jet plasma.  We also find a sizable difference in the overall level (and sign) of CP between the ordered and disordered magnetic field models.

\begin{figure}
  \setlength{\abovecaptionskip}{-6pt}
  \begin{center}
  \hspace*{-0.65cm}
    \scalebox{1.0}{\includegraphics[width=1.15\columnwidth,clip]{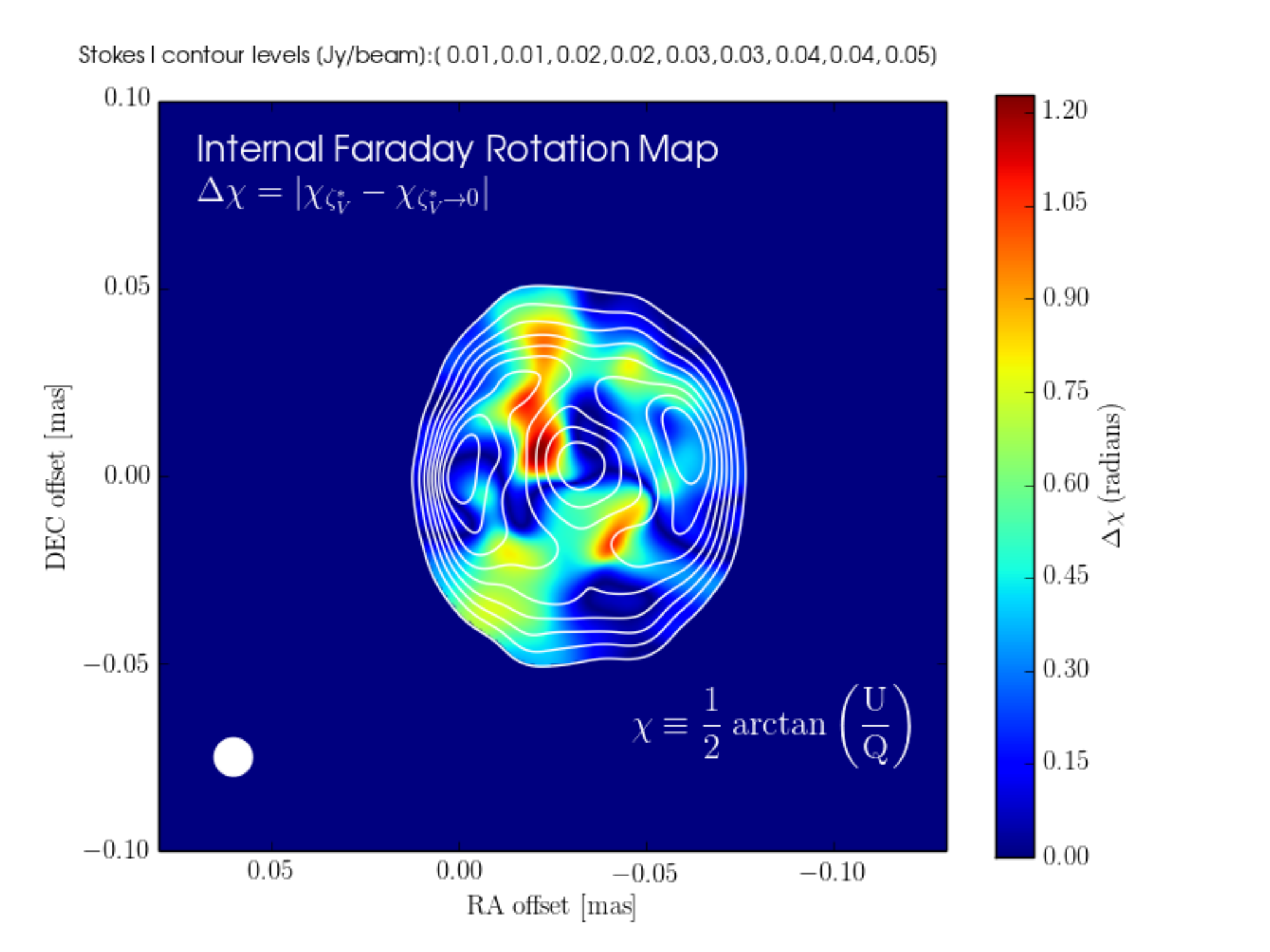}}
  \end{center}
  \caption{\label{fig7}  An internal Faraday rotation map of the TEMZ model at $\nu_{\rm{obs}} =$ 86 GHz.  At each pixel within the white Stokes I contours (see upper panel of Figure \ref{fig5}), we have computed the absolute difference in EVPA angle ($\chi$) between a ray-tracing calculation in which we include the effect of Faraday rotation ($\zeta^{*}_{V}$) and one in which we suppress Faraday rotation within the plasma ($\zeta^{*}_{V} \rightarrow 0$).}
\end{figure}

\begin{figure*}
  \setlength{\abovecaptionskip}{-6pt}
  \centering
    \includegraphics[width=0.45\linewidth]{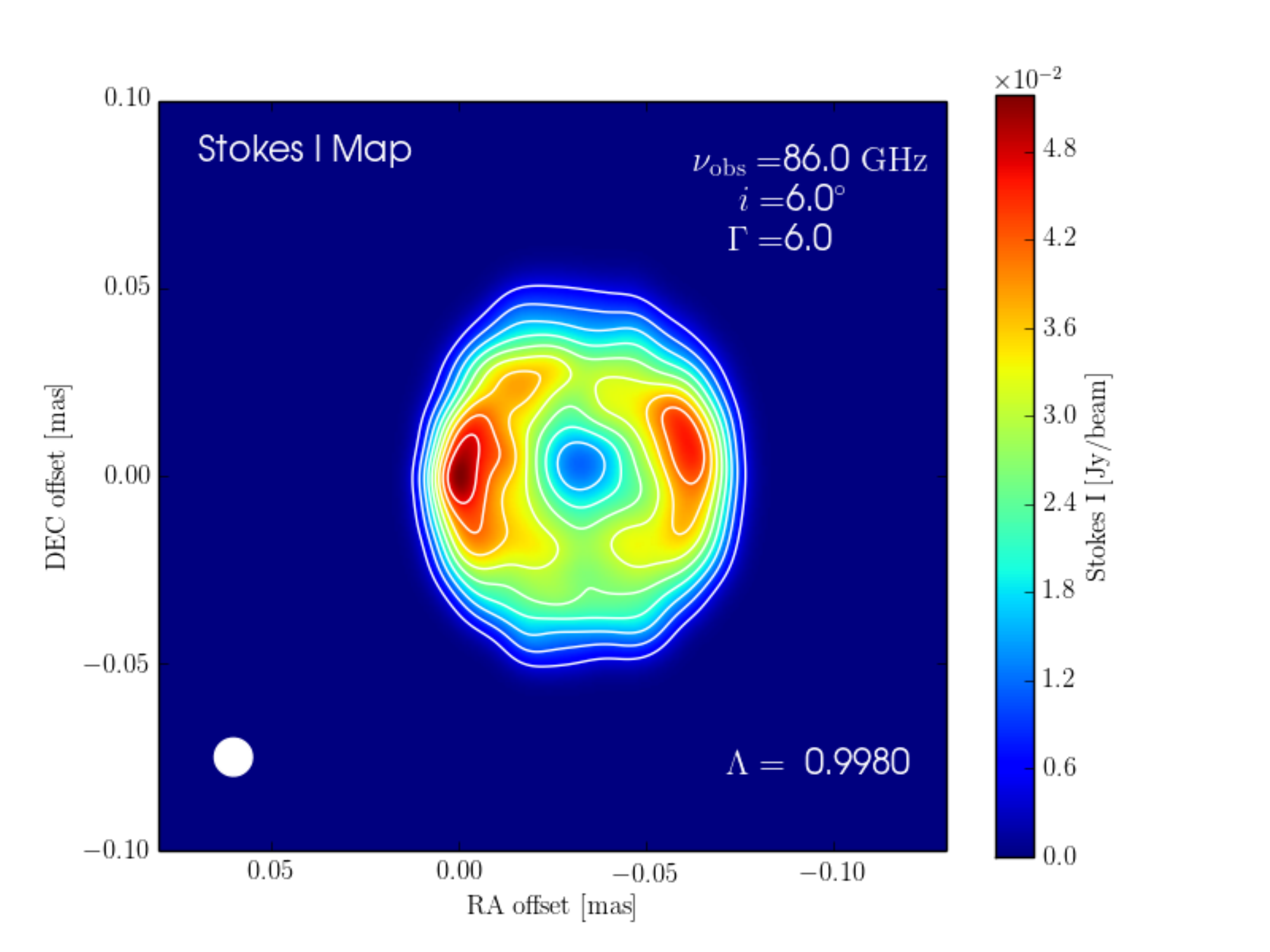}
    \includegraphics[width=0.45\linewidth]{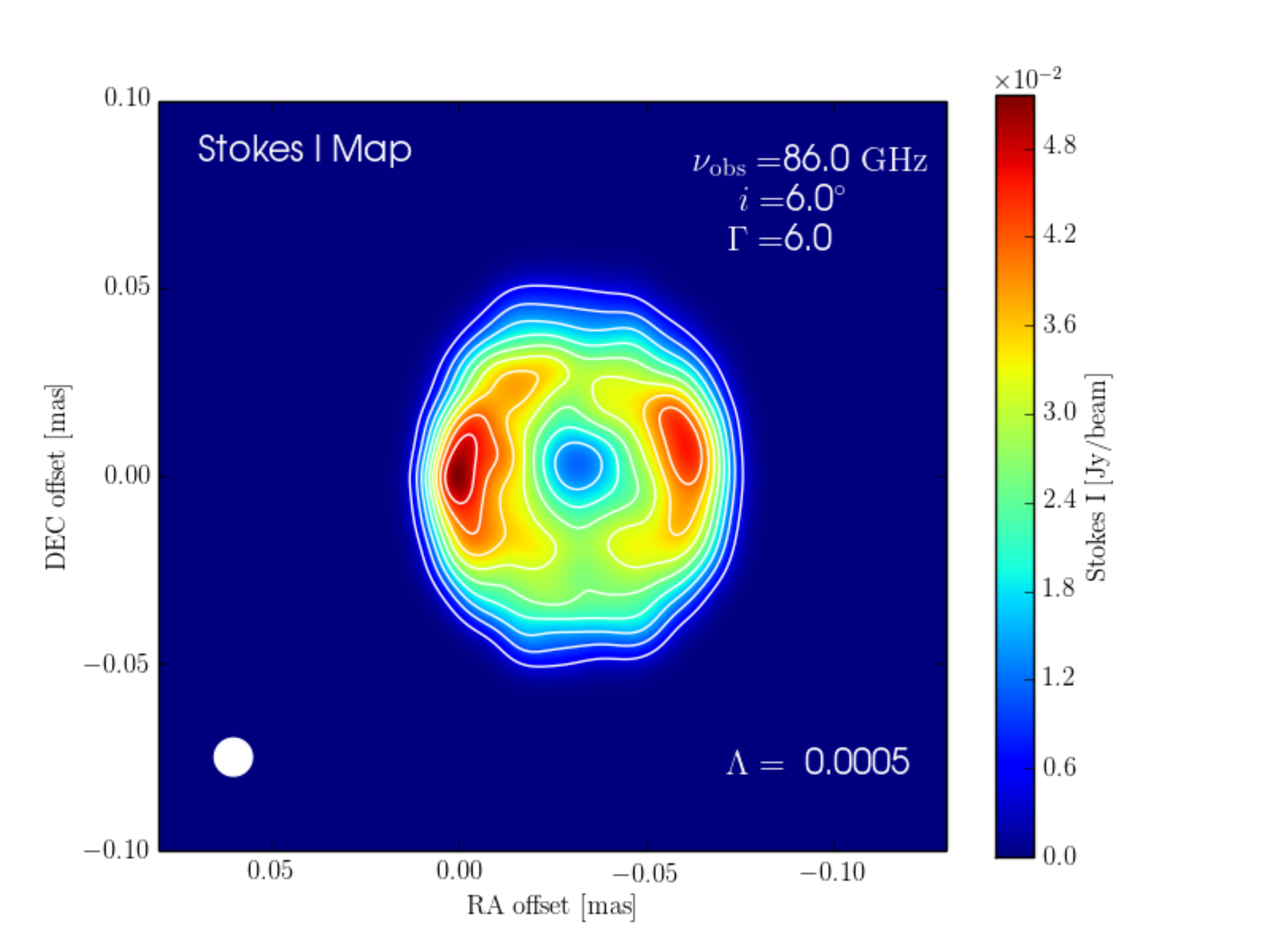}
    \includegraphics[width=0.45\linewidth]{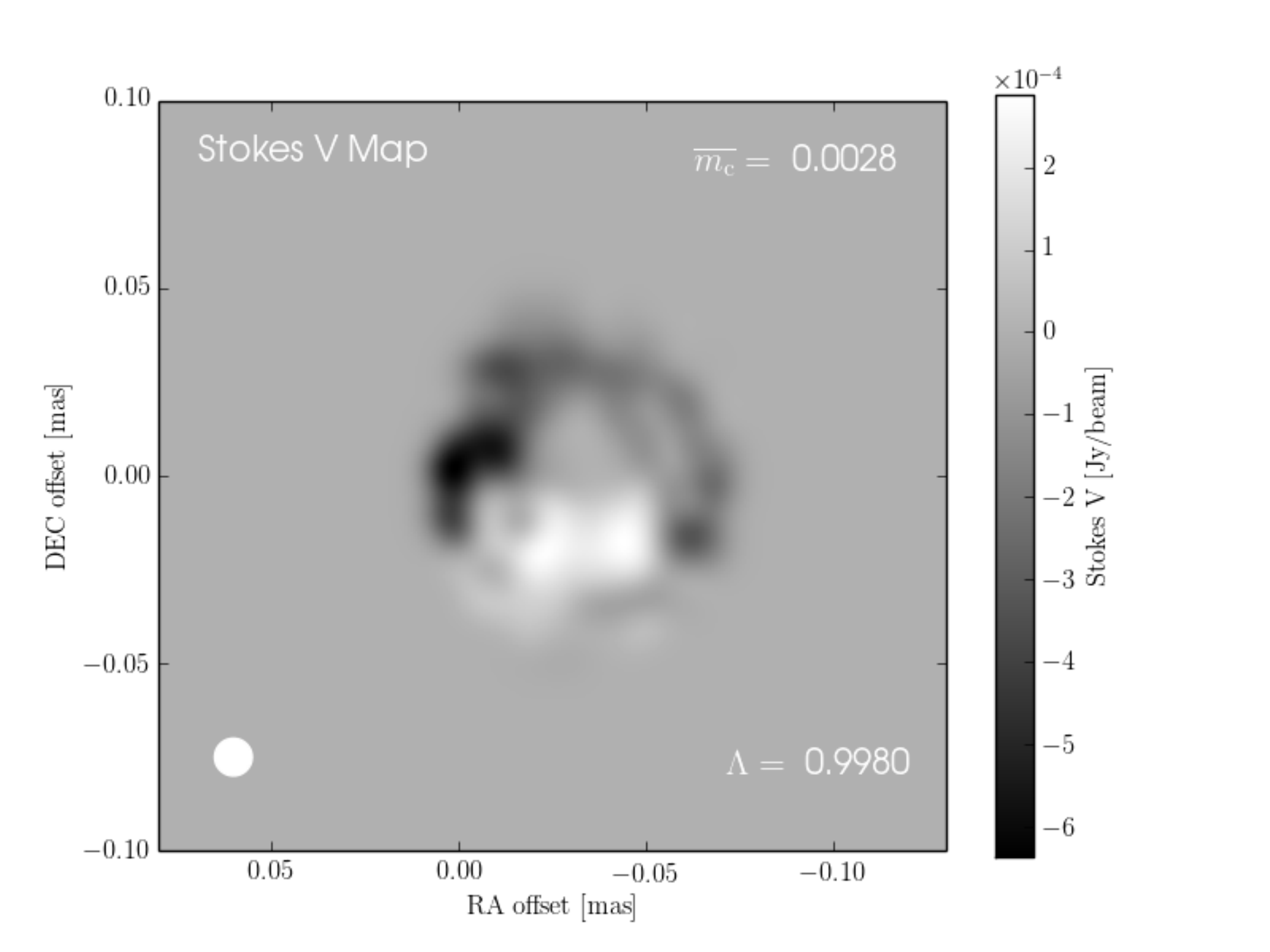}
    \includegraphics[width=0.45\linewidth]{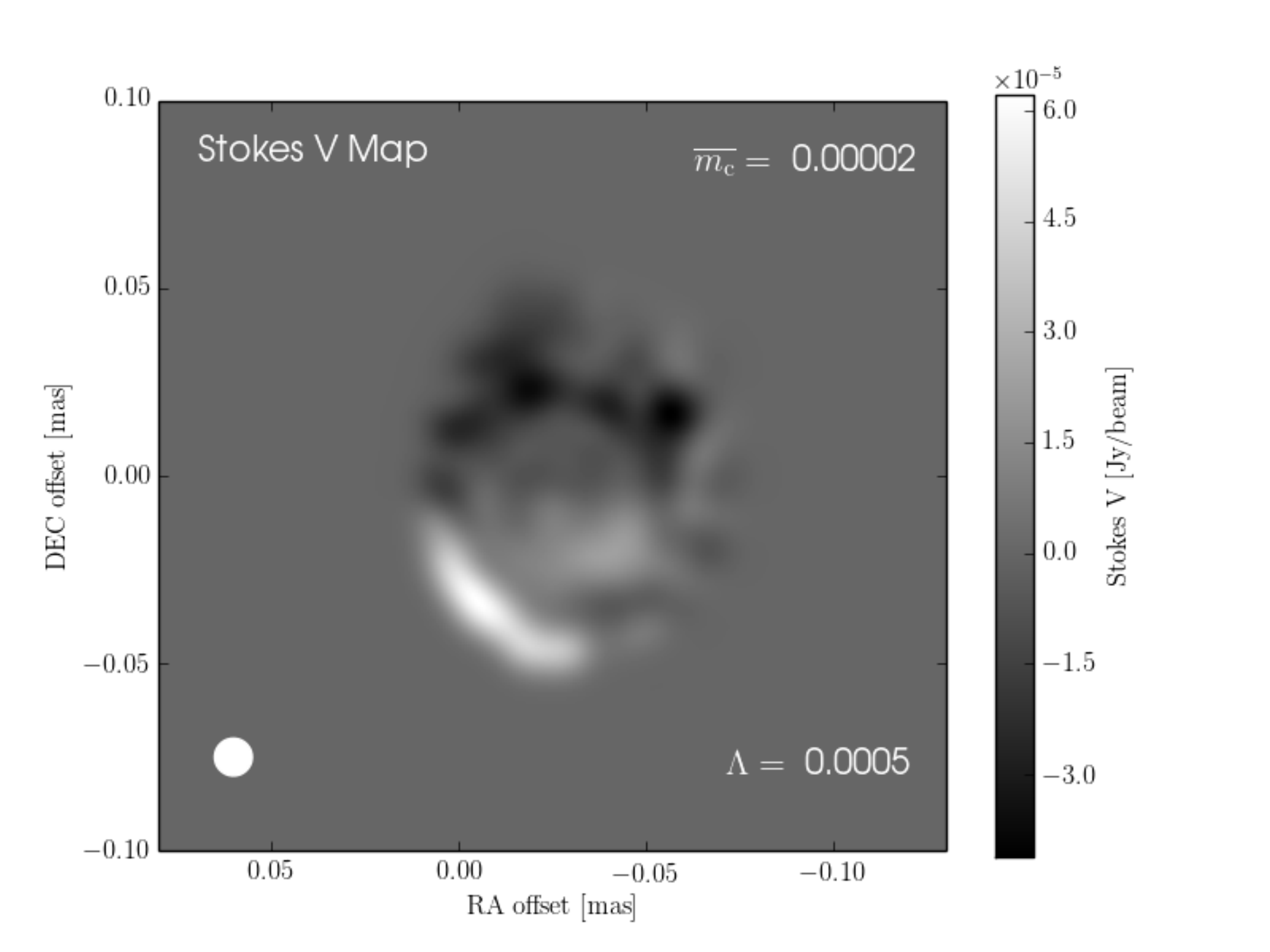}
    \vspace*{0.2cm}
  \caption{\label{fig8} \textbf{(Upper left panel)} - I map for a jet composed of ``normal'' plasma with a proton to positron ratio of 1000 ($\Lambda = 0.998$). \textbf{(Upper right panel)} - I map for a jet composed of ``pair'' plasma with a proton to positron ratio of 0.001 ($\Lambda = 0.0005$). \textbf{(Lower left panel)} - Corresponding V map for the normal plasma case.  \textbf{(Lower right panel)} -  Corresponding V map for the pair plasma case. Integrated values of $m_{\rm c}$ are listed to the upper right of the V maps. The above images have all been convolved with the circular Gaussian beam of FWHM $10 ~\mu\rm{as}$ shown in the lower left of each panel. All images are produced from the same TEMZ model at $\nu_{\rm{obs}}=$ 86 GHz.  The difference in the CP emission between the lower left and lower right panels highlights the sensitivity of CP to the underlying plasma content of the jet.}
\end{figure*}

\vfill\eject

\section{Internal Faraday Rotation}

After creating synthetic radio images of the turbulent TEMZ model at $\nu_{\rm{obs}} = 86 ~ \rm{GHz}$ (presented in \S6), we proceeded to map the amount of internal Faraday rotation occurring within the jet.  In Figure \ref{fig7}, we present a map of the degree ($\Delta \chi$):
\begin{equation}\label{eqn6}
\Delta \chi = | ~ \chi_{\zeta^{*}_{V}} - \chi_{\zeta^{*}_{V} \rightarrow 0} ~ | ~ ,
\end{equation}
by which the orientation of the observed LP angle ($\chi$) is rotated by in each pixel between a ray-tracing calculation in which we include the effects of Faraday rotation within the plasma ($\zeta^{*}_{V}$; as discussed in \S5, see Equation \ref{eqn2}) and an identical calculation in which we suppress the effects of Faraday rotation within the plasma ($\zeta^{*}_{V} \rightarrow 0$).  If the Faraday rotation is internal to the jet (as is the case here, since there is no plasma external to the jet in our computations), this degree of rotation will be:
\begin{equation}\label{eqn7}
\Delta \chi \propto \tau_{F} \propto \frac{ {\rm ln} \gamma_{\rm min} }{ \gamma_{\rm min}^{~ ~ 2 \alpha + 2} } ~\Lambda
\end{equation}
(\citealt{wardle03}).  In Equation \ref{eqn7}, $\tau_{F}$ is the Faraday rotation depth through the jet plasma, $\gamma_{\rm min}$ is the lower limit of the electron power-law energy distribution, and $\Lambda ~ (\equiv f_{l} f_{u})$ is equal to the product of the lepton charge number of the plasma ($f_{l}$) with the fraction of magnetic field that is uniform and unidirectional ($f_{u}$).   Here,
\begin{equation}\label{eqn8} 
f_{l} = \frac{(n_{-} - n_{+})}{(n_{-} + n_{+})} ~ , 
\end{equation}
where $n_{-/+}$ are the number densities of the electrons/positrons respectively.  Within each TEMZ cell the magnetic field is vector-ordered and unidirectional, implying $f_{u} \rightarrow 1$.  We explore different plasma compositions in \S10, but for this calculation we assume a normal plasma with no positrons ($n_{+} = 0 ~ \& ~f_{l} \rightarrow 1$).

\noindent As with our maps of fractional circular polarization (shown in the lower panels of Figure \ref{fig3} and \ref{fig5}), we only compute $\Delta \chi$ within a fixed limiting contour of our Stokes I map (which we arbitrarily set to 10\% of the peak I value).  We find very localized regions of relatively high internal Faraday rotation ($\Delta \chi \sim 1 ~ \rm{rad}$ at 86 GHz) within the TEMZ model.  It is within these regions that rotation driven Faraday conversion will result in the creation of CP.  This map highlights the potential of using polarimetric observations obtained on $\mu \rm{as}$ scales to constrain where in blazar jets CP is being produced through the combined relativistic processes of Faraday rotation and conversion.

\section{Plasma Composition Study}

The calculations presented above (\S4-\S9) have implicitly assumed an electron-proton plasma within the jet. Some admixture of positrons within the jet plasma should also be explored.  In order to incorporate plasma composition into our ray-tracing calculations, we have modified the normalized absorption term: $\zeta_{V}^{*}$ in our polarized radiative transfer routine (see Appendix A - Equation \ref{eqnA12}) to include an additional term: $\Lambda$, that accounts for the plasma content of the jet:
\begin{equation}\label{eqn8.5}
 \zeta_{V}^{*} = \zeta_{\alpha}^{*V} ~ ( \nu/\nu_{\rm{min}} )^{\alpha + \frac{1}{2}} ~ \frac{ \rm{ln} ~ \gamma_{\rm{min}} }{ \gamma_{\rm min} } ~ \rm{cot}( \vartheta ) ~ \Lambda ~  \left[  1 + \frac{ \alpha + 2 }{ 2 \alpha + 3 } \right] ~,
\end{equation}   
where (as discussed in \S9) $\Lambda = f_{l} f_{u}$ (\citealt{homan09}). Within an individual TEMZ cell, the local magnetic field vector is uniform and unidirectional, implying $f_{u} = 1$, while $f_{l}$ is given by Equation \ref{eqn8} (see \S9).  The number density of protons is $n_{p} = n_{-} - ~n_{+}$, since astrophysical plasmas are electrically neutral.  In our computations, a ``normal'' plasma is assigned a proton to positron ratio of 1000 ($\Lambda = 0.998$), while for a pair plasma the ratio is set at 0.001 ($\Lambda = 0.0005$).  Figure \ref{fig8} presents a comparison of the resultant Stokes I and Stokes V emission maps from snapshots of identical TEMZ models in which we assume different plasma compositions: ``normal'' plasma is shown in the left panels and ``pair'' plasma is shown in the right panels. While the I maps are indistinguishable, the differences in the morphology and level of circular polarization in the lower panels of Figure \ref{fig8} confirm that CP is indeed a sensitive probe of the plasma content of the jet. Future imaging of relativistic particle-in-cell (PIC) simulations that account for the dynamical differences between normal and pair plasma jets will be of interest.  

\newpage

\section{Spectral Polarimetry}

\begin{figure}
  \setlength{\abovecaptionskip}{-6pt}
  \begin{center}
  \hspace*{-0.65cm}
    \scalebox{1.0}{\includegraphics[width=1.15\columnwidth,clip]{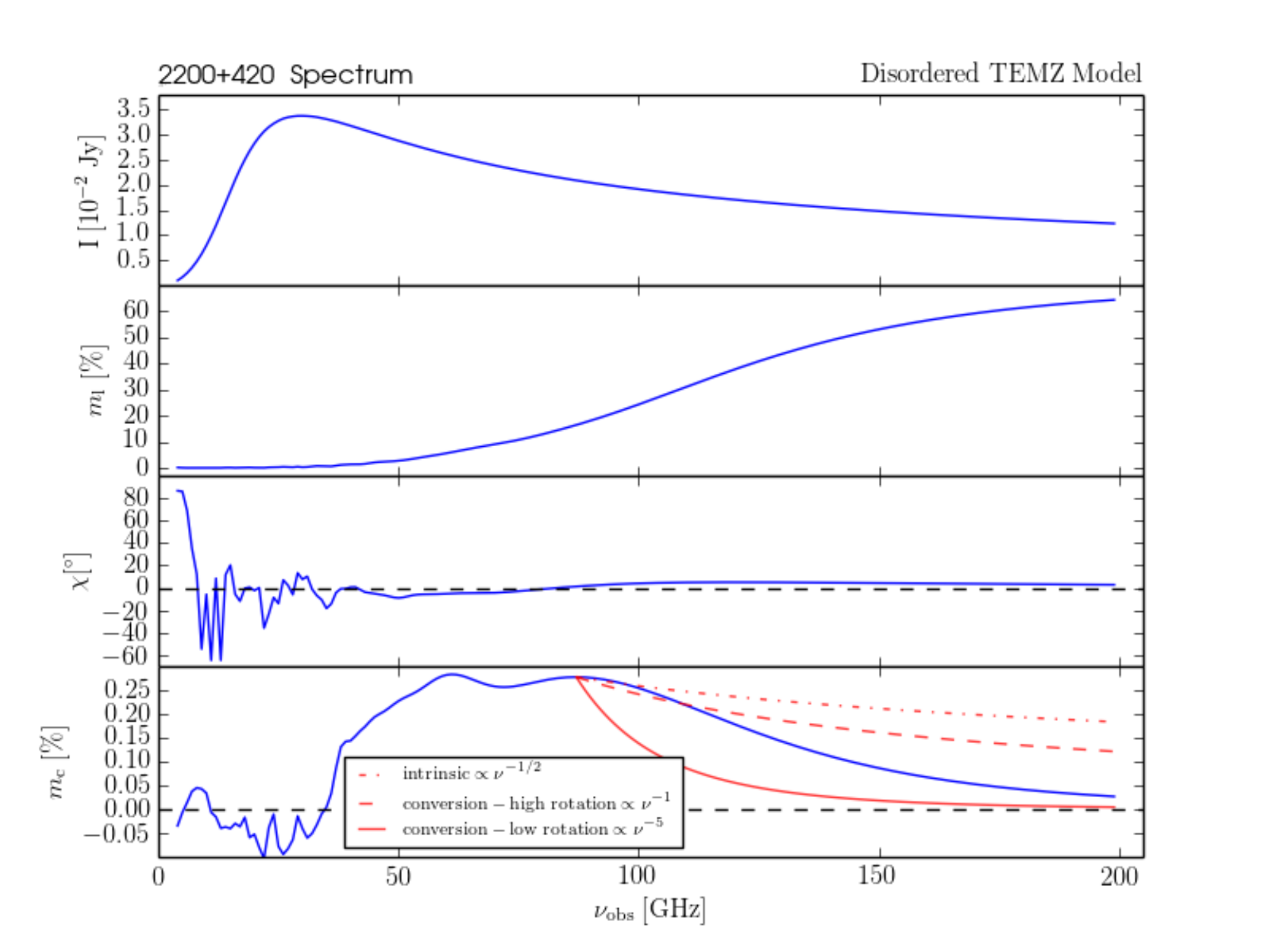}}
  \end{center}
  \caption{\label{fig9} Variations of the Stokes parameters as a function of frequency: \textbf{(Top panel)} - Integrated total intensity ($I$), \textbf{(Upper middle panel)} - Integrated fractional LP ($m_{\rm l}$), \textbf{(Lower middle panel)} - Integrated EVPA angle ($\chi$), \textbf{(Bottom panel)} - Integrated fractional CP ($m_{\rm c}$). The predicted spectral dependences of the intrinsic (dash dot), Faraday conversion in the high rotation limit (dashed), and Faraday conversion in the low rotation limit (solid) mechanisms of CP production as a function of frequency (Equations \ref{eqn9}, \ref{eqn10}, \& \ref{eqn11}) are shown as well.}
\end{figure}

As discussed in \S5, there are several distinct physical processes by which circularly polarized emission can be produced within a relativistic jet. In a homogenous plasma, CP can emanate either intrinsically ($int$) from synchrotron emission or through Faraday conversion ($con$); \cite{jones77}.  The latter process is commonly believed to be the dominant form of CP production within blazars \citep{wardle98,ruszkowski02} and can occur at both high ($\tau_{F} \gg 1$) and low ($\tau_{F} \ll 1$) Faraday rotation depths.  The spectral behavior of CP can potentially be used to discern which of these mechanisms of CP production dominates within the jet plasma. The following spectral dependancies provide useful theoretical predictions:
\begin{align}
m^{int}_{\rm C} &\propto \nu^{-1/2}\label{eqn9} \\ 
m^{con}_{\rm C ~ high ~ rotation} &\propto \nu^{-1}\label{eqn10} \\
m^{con}_{\rm C~ low ~ rotation}  &\propto \nu^{-5}\label{eqn11} ~.
\end{align}
Equation \ref{eqn9} is valid only for intrinsic optically thin emission in a uniform magnetic field, whereas Equations \ref{eqn10} and \ref{eqn11} are valid for conversion in uniform magnetic fields in the high and low rotation limits, respectively (see \citealt{jones77}, \citealt{wardle03}, and the discussion in \S5). We point out that in the standard Blandford-K\"{o}nigl jet (\citealt{blandford79}) $m_{\rm C}$ has no frequency dependence. As discussed in \cite{wardle03}, the spectral dependance of $m_{\rm c}$ comes about because the Faraday conversion ($\tau_{C}$) and rotation depths ($\tau_{F}$) are both functions of $\gamma_{\rm min}$ (see Equations \ref{eqnA11} \& \ref{eqnA12}). For an isothermal Blandford-K\"{o}nigl jet $\gamma_{\rm min}$ is constant. If, however, $\gamma_{\rm min}$ varies down the length of the jet, as it does in our model, $m_{\rm C}$ can gain a steep spectral dependence. \cite{osullivan13} have measured, with unprecedented sensitivity, both the degree and spectral behavior of CP, $m_{\rm C}(\nu)$, in the quasar PKS B2126$-$158 using the Australia Telescope Compact Array (ATCA). They found $m^{con}_{\rm C}(\nu) \sim \nu^{-3}$. They also found that the fractional LP: $m_{\rm l}(\nu)$, and the fractional CP: $m_{\rm c}(\nu)$, are anti-correlated across a frequency range of $\nu_{\rm{obs}} = 1$-10 GHz.  This spectral behavior favors the conversion ($con$) mechanism of CP production. We have created synthetic spectral polarimetric observations of the TEMZ model over a large range of frequencies (see Figure \ref{fig9}). We find that our values of integrated fractional CP, $m_{\rm c}(\nu)$, lie between the predicted curves of the high Faraday rotation depth (i.e.~circular characteristic waves) and low Faraday rotation depth (i.e.~linear characteristic waves) conversion mechanisms (Equations \ref{eqn10} \& \ref{eqn11}). This spectral behavior corresponds to elliptical characteristic waves which in an inhomogeneous plasma can result in relatively high levels of CP. We find that both $m_{\rm l}(\nu)$ and $m_{\rm c}(\nu)$ increase between $\nu_{\rm{obs}} \sim$ 30-60 GHz.  In the $\sim$100-200 GHz range, $m_{\rm l}(\nu)$ continues to increase and $m_{\rm c}(\nu)$ decreases. We do not see a clear anti-correlation between $m_{\rm l}(\nu)$ and $m_{\rm c}(\nu)$ at these higher frequencies. We emphasize, however, that our synthetic observations are on vastly different scales compared to those presented in \cite{osullivan13}. The spectra plotted in Figure \ref{fig9} (and in particular its complex low-frequency behavior) is a reflection of the inhomogeneity in the values of $\gamma_{\rm min}$ and the magnetic field strength (and orientation) through our model. In addition, our spectra are also affected by the fact that $n_{e}(\gamma)$ is changing as the electrons lose energy downstream of the shock.

\section{Summary and Conclusions}

We have found that a turbulent magnetic field partially ordered by shock compression can produce circularly polarized emission at the percent levels seen in some blazars.  We have demonstrated that Faraday conversion is indeed the dominant mechanism of CP production within the TEMZ model, in agreement with \cite{jones88} and \cite{ruszkowski02}.  We have also shown that reversals in the sign of the integrated levels of $m_{\rm C}$ are indicative of turbulent magnetic fields within the jet (as postulated by \citealt{aller03}), while constancy in the sign of $m_{\rm C}$ over many epochs can indeed be attributed to the presence of large scale ordered helical magnetic fields within the jet (as predicted by \citealt{ensslin03}).  

We have created a map of the internal Faraday rotation present within the TEMZ model at $\nu_{\rm obs} = 86 ~ \rm{GHz}$ (Figure \ref{fig7}).  The high levels of internal Faraday rotation present within our model highlights the potential of using multi-frequency rotation measures (RM) of blazar jets on $\mu$as scales to probe where in the jet Faraday conversion is occurring.  We have also confirmed (Figure \ref{fig8}) the sensitivity of CP to the underlying plasma content of the jet (as discussed in \citealt{wardle98}).  We intend to compare our synthetic emission maps to upcoming observations that will incorporate phased-ALMA and the orbiting RadioAstron antenna in mm-wave VLBI observations of blazars at 1.3 cm ($\sim 20 ~ \mu \rm{as}$ resolution).  The unprecedented sensitivity and resolution of such observational campaigns will probe where in the jets Faraday rotation and conversion is occurring.       

The next step to take with this theoretical study is to begin to explore how changes in the TEMZ model parameters affect the levels of circular polarization.  These tests will be crucial in assessing the utility of CP measurements in discerning the underlying nature of the jet plasma.  The ultimate goal of this research program is the characterization of the lepton content of the jet ($f_{l}$).  Determining this ratio is vitally important to our understanding of the effect the jet can have on its surrounding environment (``feedback'').  By combining maps of $m_{\rm C}$ and RM, we can observationally constrain the lower limit of the electron power-law energy distribution ($\gamma_{\rm min}$) and the parameter $\Lambda ~(= f_{l} f_{u})$.  We plan on compiling a set of simulations that chart a range of plausible plasma compositions ($f_{l}$) and jet magnetic field orientations ($f_{u}$).  Comparison of CP observations with this set of simulations will determine which values of $f_{l}$ and $f_{u}$ can match the data, thus probing (in a quantifiable manner) the plasma composition of a relativistic jet.  

Finally, our polarized radiative transfer scheme has been written in a robust fashion that will allow it to be used in ray-tracing through other types of numerical jet simulations (for example, particle-in cell (PIC) simulations; \citealt{nishikawa16}, or relativistic-magneto-hydrodynamic (RMHD) simulations; \citealt{marti16}).  All the routine requires is a three-dimensional distribution of magnetic fields and electron number densities from which full polarization emission maps can be generated.  

\subsection*{Acknowledgements}

Funding for this research was provided by a Canadian NSERC PGS D2 Doctoral Fellowship and by NSF grant AST-1615796.  The authors are grateful to S. G. Jorstad, J. L. G\'{o}mez, C. Casadio, I. Myserlis, E. Angelakis, and \linebreak J. D. Montgomery for helpful discussions and to the anonymous referee for a thorough review. We have made use of radmc3dPy: a python package/user-interface to the RADMC-3D code developed by A. Juhasz. Figures \ref{fig1} \& \ref{fig2} also appear in the proceedings for the conferences: ``\textit{Blazars through Sharp Multi-Wavelength Eyes}" Malaga, Spain, and ``\textit{Polarised Emission from Astrophysical Jets}" Ierapetra, Greece, both of which are published by Galaxies (an open access journal).

\newpage

\appendix
\section{\\A - Radiative Transfer of Polarized Emission}

\begin{figure}[!htbp]
\begin{center}
\includegraphics[width=0.8\textwidth]{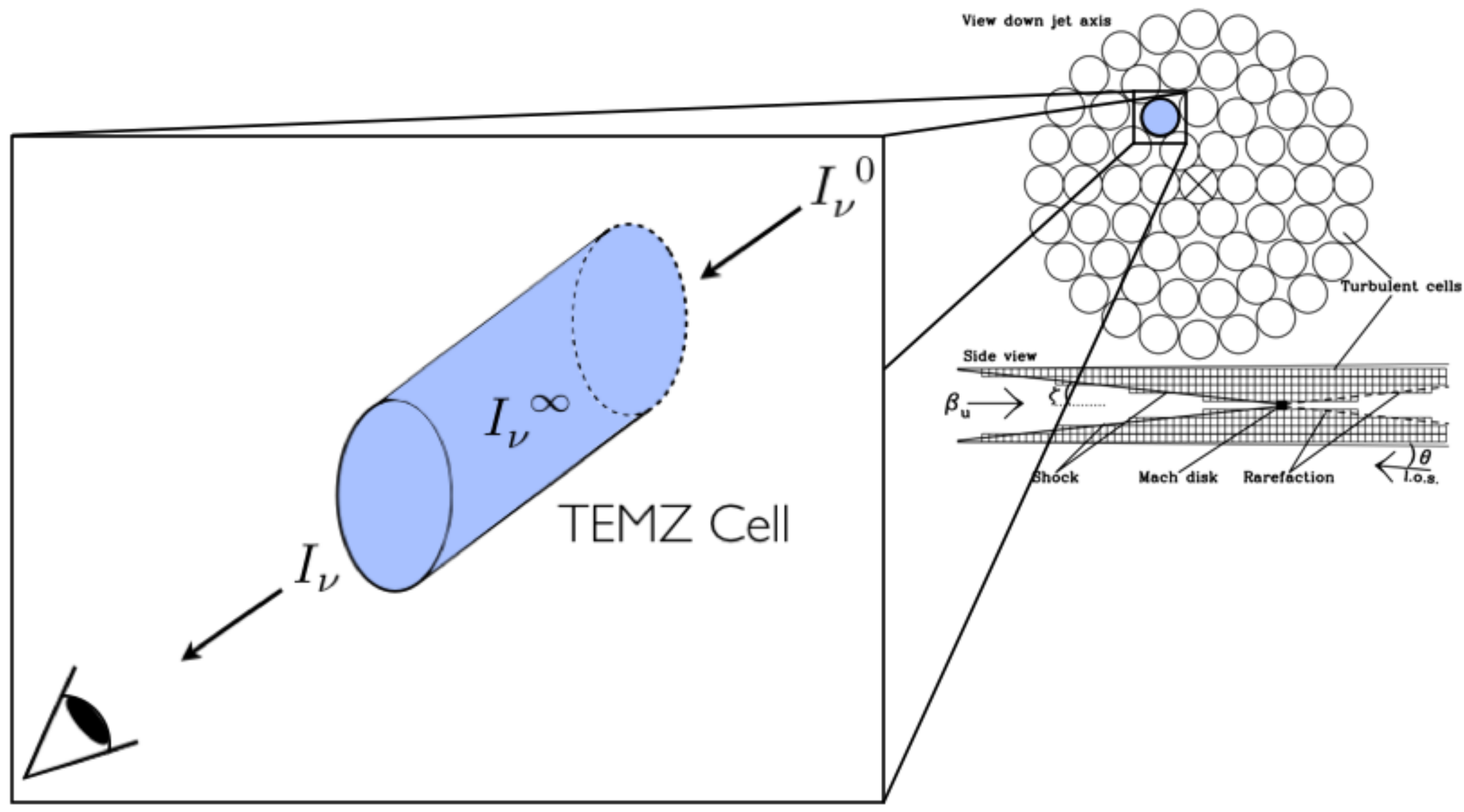}
\end{center}
\vspace{-0.3cm}
\caption{\label{fig10} Schematic representation of an individual TEMZ cell.  Emission from an adjacent cell $I_{\nu}^{~ 0}$ enters this cell and combines with the intrinsic cell emission $I_{\nu}^{~ \infty}$ to produce the resultant total intensity $I_{\nu}$ that leaves the cell along a given ray.}
\end{figure}

\subsection{Stokes I, Q, U, \& V Transfer Functions}

The full Stokes equations of polarized radiative transfer are presented in \cite{jones77}.  The matrix presented in \S3 (Equation \ref{eqn1}) can be shown to yield the following expressions for the four Stokes parameters (Equations \ref{eqnA1}, \ref{eqnA17}, \ref{eqnA18}, \& \ref{eqnA19} below) along a ray passing through a cell of magnetized plasma (see Figure \ref{fig10}).  These solutions involve an arbitrary rotation of coordinates to a reference frame in the plasma in which the emission and absorption coefficients of Stokes U are $\eta_{\nu}^{~ U} = \kappa_{U} = \kappa_{U}^{*} = 0$.  This rotation, discussed in \cite{jones77}, makes the mathematics more tractable.  In particular, the observed total intensity ($I_{\nu}$) from a given cell is a ``blend'' of intrinsic cell emission ($I_{\nu}^{~ \infty}$) with external emission ($I_{\nu}^{~ 0}$) passing into the cell from an adjacent cell along a particular sight-line: 

\begin{align}\label{eqnA1}
I_{\nu} = I_{\nu}^{~ \infty} + \rm{e}^{-\tau} \Biggl\{ ~\; \rm{cosh}( \chi \tau )& \left[\frac{1}{2}( 1 + q^{2} + v^{2} )(\: I_{\nu}^{~ 0} - I_{\nu}^{ ~ \infty } \:) + (\;q \times v\:)( U_{\nu}^{~ 0} - U_{\nu}^{~ \infty} )\right] \nonumber \\ -\rm{sinh}( \chi \tau )& \: \bigg[ \quad\!(\quad\;\: q \cdot k \quad\;\:)(Q_{\nu}^{~ 0} - Q_{\nu}^{~ \infty})\! + (\;\, v \cdot k \;\,)( V_{\nu}^{~ 0} - V_{\nu}^{~\infty} )\bigg] \nonumber \\ +\rm{cos}( \chi_{*} \tau )& \left[\frac{1}{2}( 1 -q^{2} - v^{2} )(\: I_{\nu}^{~0} - I_{\nu}^{~\infty} \:) - (\;q \times v\:)( U_{\nu}^{~ 0} - U_{\nu}^{~ \infty} )\right]  \nonumber \\ -\rm{sin}( \chi_{*} \tau )& \:\bigg[ \quad\!(\quad\;\:q \times k\quad\!)(Q_{\nu}^{~ 0} - Q_{\nu}^{~ \infty})\! + (\; v \times k \;)( V_{\nu}^{~ 0} - V_{\nu}^{~ \infty} )\bigg] ~\; \Biggr\} ~ .
\end{align}

\noindent The optical depth of the cell is denoted by $\tau$ and is a function of the cell's opacity $\kappa$ integrated along the ray's path length $l$ through the cell: $\tau = \int \kappa ~ \rm{d}l$.  The quantities $\chi$ and $\chi_{*}$, as well as $q$, $v$, and $k$, are defined in \cite{jones77} and pertain to the effects of Faraday rotation and conversion acting on both the radiation passing through the cell and the radiation intrinsic to the cell.  This implies that each cell in the TEMZ computational grid is an ``internal'' Faraday screen for radiation emanating from adjacent cells along any given sight-line.  As $\tau \rightarrow \infty$, Equation \ref{eqnA1} yields: $I_{\nu} \rightarrow I_{\nu}^{~ \infty}$ (i.e., in the optically thick case the observed intensity is simply equal to the source function intrinsic to the cell), and as $\tau \rightarrow 0$, Equation \ref{eqnA1} yields: $I_{\nu} \rightarrow I_{\nu}^{~ 0}$ (i.e., in the optically thin case the observed intensity is simply equal to the unmodified emission passing through the cell). The solutions to the other three Stokes parameters (analogous to Equation \ref{eqnA1}) are given by the following expressions:

\begin{align}\label{eqnA17}
Q_{\nu} = Q_{\nu}^{~ \infty} + \rm{e}^{-\tau} \Biggl\{ ~\; \rm{cosh}( \chi \tau )& \left[\frac{1}{2}( 1 + q^{2} - v^{2} )( Q_{\nu}^{~ 0} - Q_{\nu}^{ ~ \infty } ) + (\;q \cdot v\;)( V_{\nu}^{~ 0} - V_{\nu}^{~ \infty} )\right] \nonumber \\ -\rm{sinh}( \chi \tau )& \: \bigg[ \quad\!(\quad\;\: q \cdot k \quad\;\:)( \; I_{\nu}^{~ 0} - I_{\nu}^{~ \infty} \, ) - (v \times k )( U_{\nu}^{~ 0} - U_{\nu}^{~\infty} )\bigg] \nonumber \\ +\rm{cos}( \chi_{*} \tau )& \left[\frac{1}{2}( 1 -q^{2} + v^{2} )( Q_{\nu}^{~0} - Q_{\nu}^{~\infty} ) - (\;q \cdot v\;)( V_{\nu}^{~ 0} - V_{\nu}^{~ \infty} )\right]  \nonumber \\ -\rm{sin}( \chi_{*} \tau )& \:\bigg[ \quad\!(\quad\;\:q \times k\quad\!)( \; I_{\nu}^{~ 0} - I_{\nu}^{~ \infty} \; ) + (\; v \cdot k \;)( U_{\nu}^{~ 0} - U_{\nu}^{~ \infty} )\bigg] ~\; \Biggr\}
\end{align}

\begin{align}\label{eqnA18}
U_{\nu} = U_{\nu}^{~ \infty} + \rm{e}^{-\tau} \Biggl\{ ~\; \rm{cosh}( \chi \tau )& \left[-( q \times v )(\; I_{\nu}^{~ 0} - I_{\nu}^{ ~ \infty } \;) + \frac{1}{2}(1 - q^{2} - v^{2})( U_{\nu}^{~ 0} - U_{\nu}^{~ \infty} )\right] \nonumber \\ -\rm{sinh}( \chi \tau )& \: \bigg[ \quad\!(\, v \cdot k \,)(Q_{\nu}^{~ 0} - Q_{\nu}^{~ \infty}) - \;\;\,(\quad\; q \times k \quad\;)( V_{\nu}^{~ 0} - V_{\nu}^{~\infty} )\bigg] \nonumber \\ +\rm{cos}( \chi_{*} \tau )& \left[\quad\!( q \times v )(\; I_{\nu}^{~0} - I_{\nu}^{~\infty} \;) + \frac{1}{2}(1+q^{2}+v^{2})( U_{\nu}^{~ 0} - U_{\nu}^{~ \infty} )\right]  \nonumber \\ -\rm{sin}( \chi_{*} \tau )& \,\bigg[\!-\!(\, v \cdot k \,)( Q_{\nu}^{~ 0} - Q_{\nu}^{~ \infty} ) + \;\;\,(\quad\;\: q \cdot k \quad\;\:)( V_{\nu}^{~ 0} - V_{\nu}^{~ \infty} )\bigg] ~\; \Biggr\}
\end{align}

\begin{align}\label{eqnA19}
V_{\nu} = V_{\nu}^{~ \infty} + \rm{e}^{-\tau} \Biggl\{ ~\; \rm{cosh}( \chi \tau )& \left[\quad\!(\; q \cdot v \;)( Q_{\nu}^{~ 0} - Q_{\nu}^{ ~ \infty } ) + \frac{1}{2}( 1 - q^{2} + v^{2} )( V_{\nu}^{~ 0} - V_{\nu}^{~ \infty} )\right] \nonumber \\ -\rm{sinh}( \chi \tau )& \: \bigg[ \quad\!(\; v \cdot k \;)(\; I_{\nu}^{~ 0} - I_{\nu}^{~ \infty} \;) + \;\;\,(\quad\; q \times k \quad\;)( U_{\nu}^{~ 0} - U_{\nu}^{~\infty} )\bigg] \nonumber \\ +\rm{cos}( \chi_{*} \tau )& \left[-(\; q \cdot v \;\,)( Q_{\nu}^{~0} - Q_{\nu}^{~\infty} ) + \frac{1}{2}( 1 +q^{2} - v^{2} )( V_{\nu}^{~ 0} - V_{\nu}^{~ \infty} )\right]  \nonumber \\ -\rm{sin}( \chi_{*} \tau )& \:\bigg[ \quad\!( v \times k )( \; I_{\nu}^{~ 0} - I_{\nu}^{~ \infty} \; ) - \;\;\,(\quad\;\:\, q \cdot k \quad\;\:)( U_{\nu}^{~ 0} - U_{\nu}^{~ \infty} )\bigg] ~\; \Biggr\} ~.
\end{align}

\noindent Similar to Equation \ref{eqnA1}, as $\tau \rightarrow \infty$, Equations (\ref{eqnA17}-\ref{eqnA19}) yield: $Q_{\nu},U_{\nu},V_{\nu} \rightarrow Q_{\nu}^{~ \infty},U_{\nu}^{~ \infty},V_{\nu}^{~ \infty}$, respectively, and as $\tau \rightarrow 0$, Equations (\ref{eqnA17}-\ref{eqnA19}) yield: $Q_{\nu},U_{\nu},V_{\nu} \rightarrow Q_{\nu}^{~ 0},U_{\nu}^{~ 0},V_{\nu}^{~ 0}$.

\subsection{Optical Depth}
\label{O}

\begin{figure}[!htbp]
\begin{center}
\includegraphics[width=0.3\textwidth]{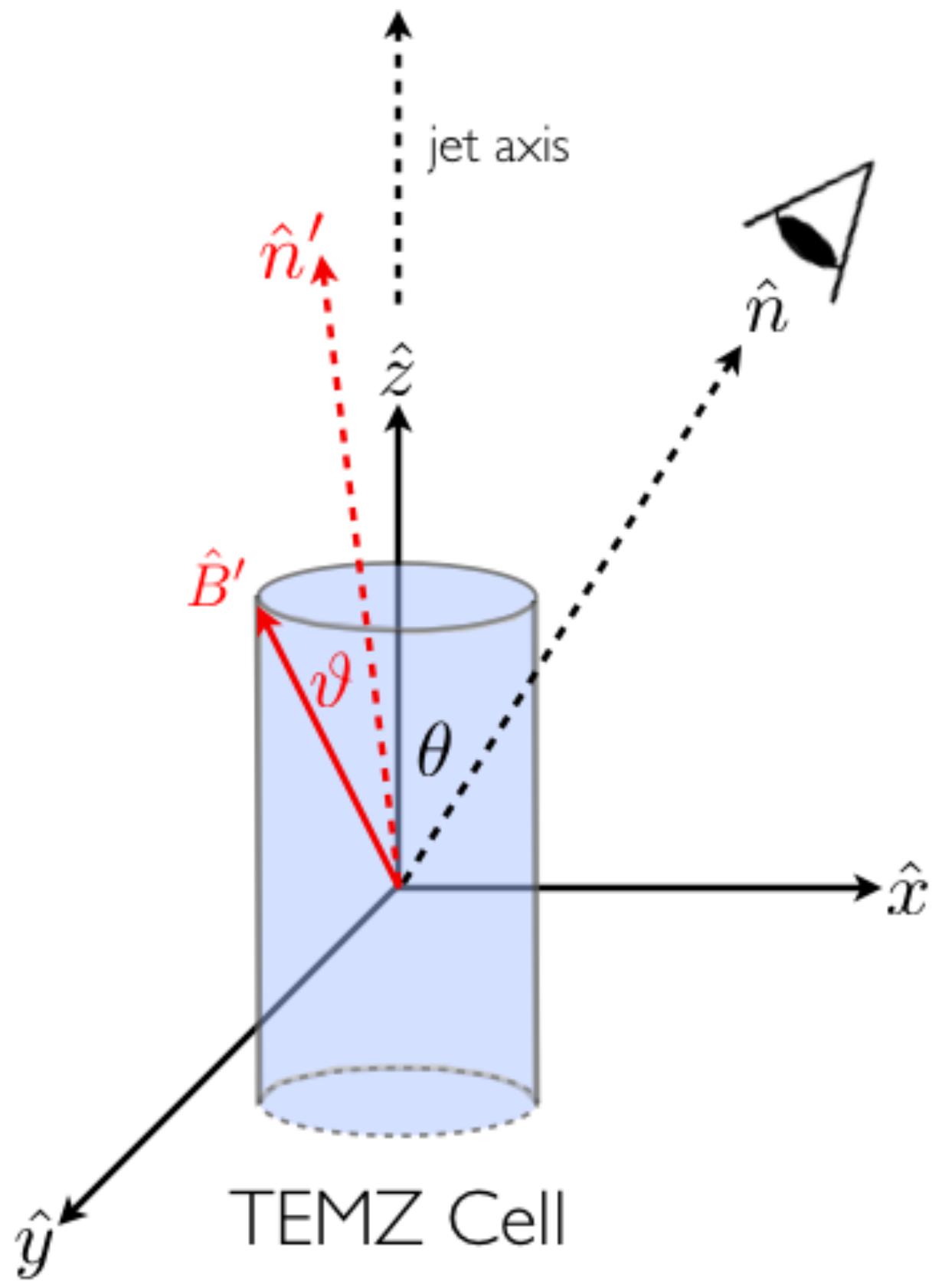}
\end{center}
\vspace{-0.3cm}
\caption{\label{fig11} A schematic representation of the effect of relativistic aberration on the angle $\vartheta$ between the line of sight and the local magnetic field vector within the co-moving frame of the plasma cell.  The jet axis is set to $\hat{z}$ and, in the observer's frame, the angle of inclination of the jet to the line of sight is $\theta$.  Owing to relativistic aberration, the sight-line is aberrated from $\hat{n}$ to $\hat{n}^{\prime}$ in the co-moving frame of the plasma.  In this figure a prime (in red) denotes the co-moving frame of the plasma.}
\end{figure}

Following \cite{jones77} (their equations C4, C6, \& C17 for the transfer coefficients of the synchrotron process), the optical depth through a TEMZ cell is evaluated as follows: 

\begin{align}\label{eqnA2}
\tau &= \int \kappa ~ \rm{d}l  \nonumber \\
        &= \int \kappa_{\alpha} ~ \kappa_{\perp} ~ ( \nu_{B \perp}/\nu )^{\alpha + 5/2} ~ \rm{d}l \nonumber \\
        &= \int \kappa_{\alpha} ~ ( r_{e}c ) ~ \nu_{B \perp}^{ ~ ~ ~ -1} ~ [ ~ 4 \pi g(\vartheta) ~ ] ~ [ ~ n_{0} ~ ] ~ ( \nu_{B \perp}/\nu )^{\alpha + 5/2} ~ \rm{d}l \nonumber \\
        &= \kappa_{\alpha} ~ ( r_{e}c ) ~ \nu_{B \perp}^{ ~ ~ ~ -1} ~ [ ~ 4 \pi g(\vartheta) ~ ] ~ [ ~ n_{0} ~ ] ~ ( \nu_{B \perp}/\nu )^{\alpha + 5/2} ~ l ~ ,
\end{align}

\noindent where $l$ is the path length through the TEMZ cell (evaluated by RADMC-3D) and $n_{0}$ is the electron number density within the cell; the plasma is homogeneous within each cell.  The characteristic length scale of a TEMZ cell for the simulations presented in this paper was set to $0.004 ~ \rm{pc}$.  The following power-law energy distribution is set in each cell: $n(\gamma) = \int n_{o} \gamma^{-s} \rm{d} \gamma$, where the energy of a given electron is $E = \gamma m_{e} c^{2}$ and $s$ is a power-law index (related to the optically thin spectral index $\alpha$ by $s=2\alpha + 1$).  This power-law is assumed valid over the energy range $\gamma_{\rm{min}}$ to $\gamma_{\rm{max}}$, as discussed in \S4.  The parameter $\kappa_{ \alpha }$ is a physical constant (of order unity) that is tabulated in \cite{jones77} and depends on the value of the spectral index $\alpha$.  The quantity $\nu_{B \perp}$ is given by:

\begin{equation}\label{eqnA3} 
\nu_{B \perp} = \frac{ e B ~ \rm{sin} \vartheta }{ 2 \pi ~ m_{e} c } ~ ,  
\end{equation}

\noindent where $\vartheta$ is the angle that the line of sight makes with respect to the local magnetic field vector in the co-moving frame of the plasma cell (see Figure \ref{fig11}).  

\subsection{Relativistic Aberration}
\label{Bonn}

When computing $\vartheta$, one needs to account for the effects of relativistic aberration; in particular, the following Lorentz transformation maps our sight-line unit vector $\hat{n}$ in the observer's frame to the corresponding sight-line unit vector $\hat{n}^{\prime}$ in the co-moving frame of the plasma:

\begin{equation}\label{eqnA4} 
\hat{n}^{\prime} = \frac{ \hat{n} + \Gamma \vec{ \beta } ~ [ ~ \frac{ \Gamma }{ \Gamma + 1 } \beta \rm{cos}( \theta ) - 1 ~  ] }{ \Gamma ~ [ ~ 1 - \beta \rm{cos}( \theta ) ~ ] } ~ ,  
\end{equation}

\noindent \citep*{lyutikov05}, where $\Gamma$ is the bulk Lorentz factor of the jet flow and $\vec{\beta} \equiv \vec{v}_{\rm{ jet }}/c$.  As illustrated in Figure \ref{fig11}, the jet velocity is assumed to be along the $\hat{z}$-axis and therefore $\vec{ \beta } = \{~ 0, ~ 0, ~ \beta ~\}$, where $\beta = \sqrt{ 1 - \frac{ 1 }{ \Gamma^{2} } }$.  From Figure \ref{fig11} it also follows that: $\hat{n} = \{~  \rm{sin}( \theta ), ~ 0, ~ \rm{cos}( \theta ) ~\}$, where $\theta$ is the angle of inclination of the jet to the line of sight in the observer's frame.  Recalling the definition of the relativistic Doppler boosting factor: $\delta \equiv \frac{ 1 }{ \Gamma(~ 1 - \beta \rm{cos}( \theta ) ~) }$, Equation \ref{eqnA4} can be rewritten in component form as:

\begin{align}\label{eqnA5}
\hat{n}^{\prime} &= \delta ~ \{~ \rm{sin}( \theta ), ~ 0, ~ \rm{cos}( \theta ) +  \Gamma \beta ~ \left[ ~ \frac{ \Gamma }{ \Gamma + 1 } \beta \rm{cos}( \theta ) - 1 ~  \right] ~ \} \nonumber \\
			&= \delta ~ \{~ \rm{sin}( \theta ), ~ 0, ~ \Gamma [ ~ \rm{cos}( \theta ) - \beta ~ ]  ~ \} ~ ,
\end{align} 

\noindent after combining terms and simplifying.  In the co-moving frame of the plasma, the local magnetic field vector is given by $\hat{B}^{\prime} = ( ~ B_{x}, ~ B_{y}, ~ B_{z} ~ )$.  It then follows that:

\begin{equation}\label{eqnA6}
\rm{cos}( \vartheta ) = \hat{n}^{\prime} \cdot \hat{B}^{\prime} \rightarrow \vartheta = \rm{cos}^{-1} \Biggl\{ \delta ~ \rm{sin}( \theta ) ~ B_{x} + \delta ~ \Gamma [  ~ \rm{cos}(\theta) - \beta ~ ] ~ B_{z} \Biggl\} ~ .
\end{equation}  

\noindent In Equation \ref{eqnA2}, the term $g(\vartheta)$ is the pitch angle distribution of the electrons within the plasma.  Here we have assumed an isotropic pitch angle distribution, $g(\vartheta) = \frac{1}{2} ~ \rm{sin}( \vartheta )$.

\subsection{Faraday Rotation and Conversion Depths}   

As discussed in \S4, the Faraday rotation ($\tau_{F}$) and conversion ($\tau_{C}$) depths through a TEMZ cell are given by $\tau_{F} = |\zeta_{V}^{*}| ~ \tau$ and $\tau_{C} = |\zeta_{Q}^{*}| ~ \tau$ respectively. The expressions $\zeta_{V}^{*}$ and $\zeta_{Q}^{*}$ represent normalized plasma absorption coefficients: $\zeta_{(Q,V)}^{*} \equiv \kappa^{*}_{(Q,V)}/\kappa_{I}$ (i.e., the $\kappa$'s in the matrix shown in \S3) and are defined as follows: 

\begin{align}
              \zeta_{Q}^{*} &=-\zeta_{\alpha}^{*Q} ~ ( \nu/\nu_{\rm{min}} )^{\alpha - \frac{1}{2}} ~ \left\{ \left[  1 - \left( \frac{\nu_{\rm{min}}}{\nu} \right)^{ \alpha - \frac{1}{2} } \right] \left( \alpha - \frac{1}{2} \right)^{-1} \right\} ~ \rm{for} ~ \alpha > \frac{1}{2}\label{eqnA11}  \\
              \zeta_{V}^{*} &=\; \; \: \zeta_{\alpha}^{*V} ~ ( \nu/\nu_{\rm{min}} )^{\alpha + \frac{1}{2}} ~ \frac{ \rm{ln} ~ \gamma_{\rm{min}} }{ \gamma_{\rm min} } ~ \rm{cot}( \vartheta )  \left[  1 + \frac{ \alpha + 2 }{ 2 \alpha + 3 } \right]\label{eqnA12} ,
\end{align}

\noindent where $\zeta_{\alpha}^{*Q}$ and $\zeta_{\alpha}^{*V}$ are physical constants (of order unity) that are tabulated in \cite{jones77} and depend on the value of the spectral index $\alpha$.  The quantity $\nu_{\rm{min}} \equiv \gamma_{\rm{min}}^{2} \nu_{B \perp}$.  In all of the computations presented in this paper, we have assumed a constant optically thin spectral index of $\alpha = 0.65$ ($s=2.3$). Equations \ref{eqnA11} \& \ref{eqnA12} are only applicable provided: $\nu_{\rm{obs}} > \nu_{\rm{min}}$. We have confirmed numerically that this criterion is met in all cells in all of our ray-trace calculations presented in this paper.

\subsection{Intrinsic Circular Polarization as a Test of the Algorithm}
\label{Wardle}

\begin{figure}[!htbp]
\begin{center}
\includegraphics[width=0.6\textwidth]{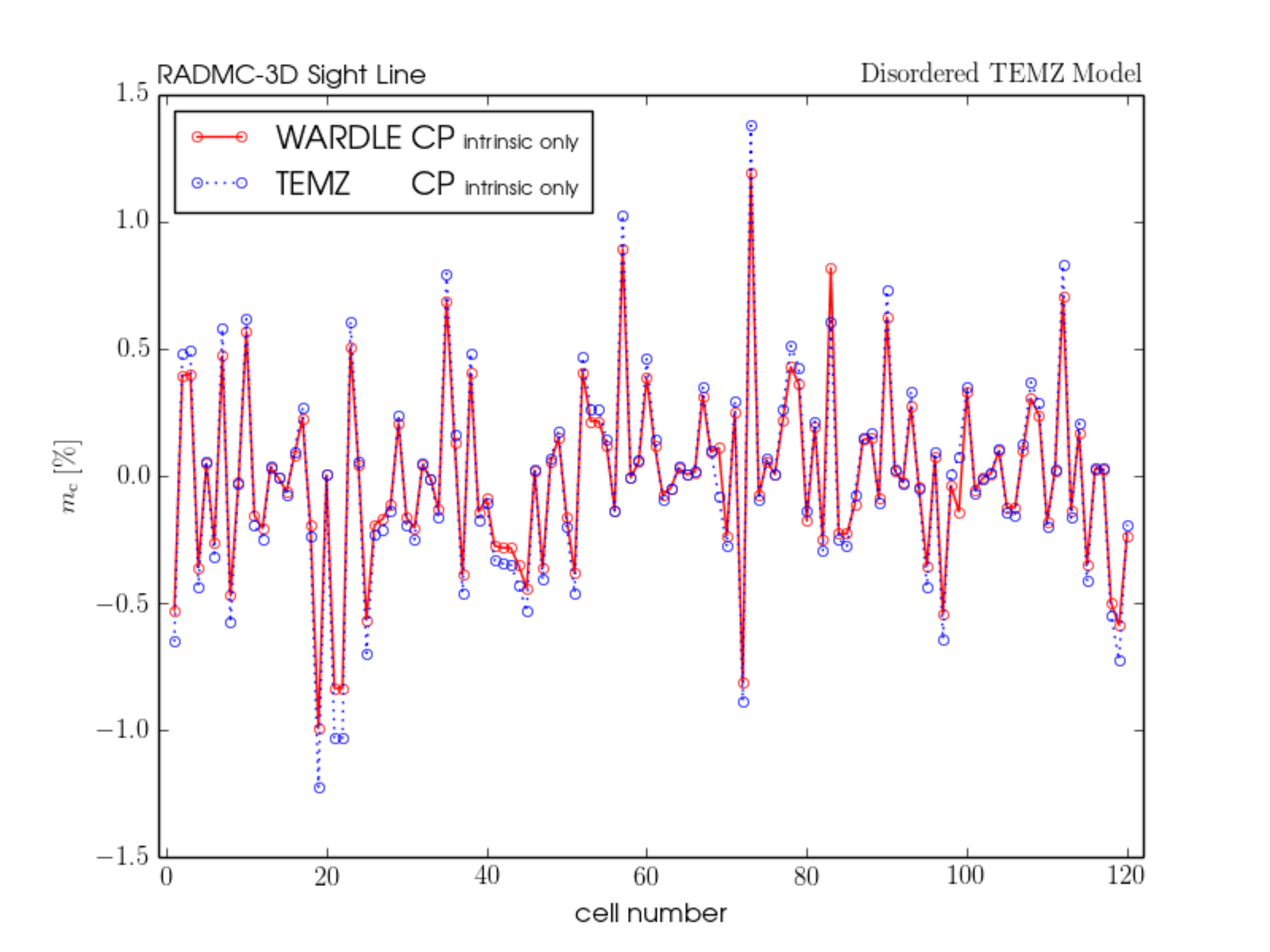}
\end{center}
\vspace{-0.4cm}
\caption{\label{fig12} Comparison of values of fractional CP ($m_{\rm c} = -V/I$) computed within individual plasma cells of the TEMZ grid along an arbitrarily selected RADMC-3D sight-line.  The radiative transfer starts at the back of the jet (cell 0) and progresses to the front  (cell 120), resulting in the creation of one pixel in our image maps.  The red circles correspond to values computed with Equation (\ref{eqnA20}), an analytic expression for $m_{\rm c}$ taken from \cite{wardle03}. The blue circles are computed by dividing Equation (\ref{eqnA19}) by Equation (\ref{eqnA1}), taken from \cite{jones77}, where we have set $I_{\nu}^{~ 0} = Q_{\nu}^{~0} = U_{\nu}^{~0} = V_{\nu}^{~ 0} = 0$ to ensure that the emission is intrinsic.  The cell-to-cell agreement between these two quantities along this sight-line represents a successful test of the numerical code.}
\end{figure}

Equations (\ref{eqnA1}-\ref{eqnA19}) have been incorporated into a numerical algorithm to solve for $I_{\nu}$, $Q_{\nu}$, $U_{\nu}$, and $V_{\nu}$ along rays passing through the TEMZ grid.  This numerical algorithm has been embedded into the ray-tracing code RADMC-3D. Each subroutine within this algorithm has been carefully vetted by comparing computed values to analytic calculations.  As a further test of this algorithm, we compute the intrinsic fractional circular polarization ($m_{\rm c} = -V/I$) within each cell along a particular sight-line by setting $I_{\nu}^{~ 0} = Q_{\nu}^{~0} = U_{\nu}^{~0} = V_{\nu}^{~ 0} = 0$ while ray-tracing with RADMC-3D.  We compare our computed values (Equation \ref{eqnA19} divided by \ref{eqnA1}) to an analytic expression for intrinsic $m_{\rm c}$ presented in \cite{wardle03}:
\begin{equation}\label{eqnA20}
m_{\rm c} = -\frac{ V }{ I } = \epsilon_{\alpha}^{~V} \left( \frac{\nu_{B \perp}}{\nu} \right)^{1/2} \rm{cot} \vartheta ~ .
\end{equation}  
The results of this comparison along a random sight-line through the TEMZ grid is illustrated in Figure \ref{fig12}.  The excellent agreement between the numerical and analytical calculations implies that Equations (\ref{eqnA1}-\ref{eqnA19}) have been incorporated correctly into our numerical algorithm. 

\newpage

\section{\\B - Global VLBI Array Resolution Images}

\begin{figure*}[!htbp]
\centering
\includegraphics[width=0.49\linewidth]{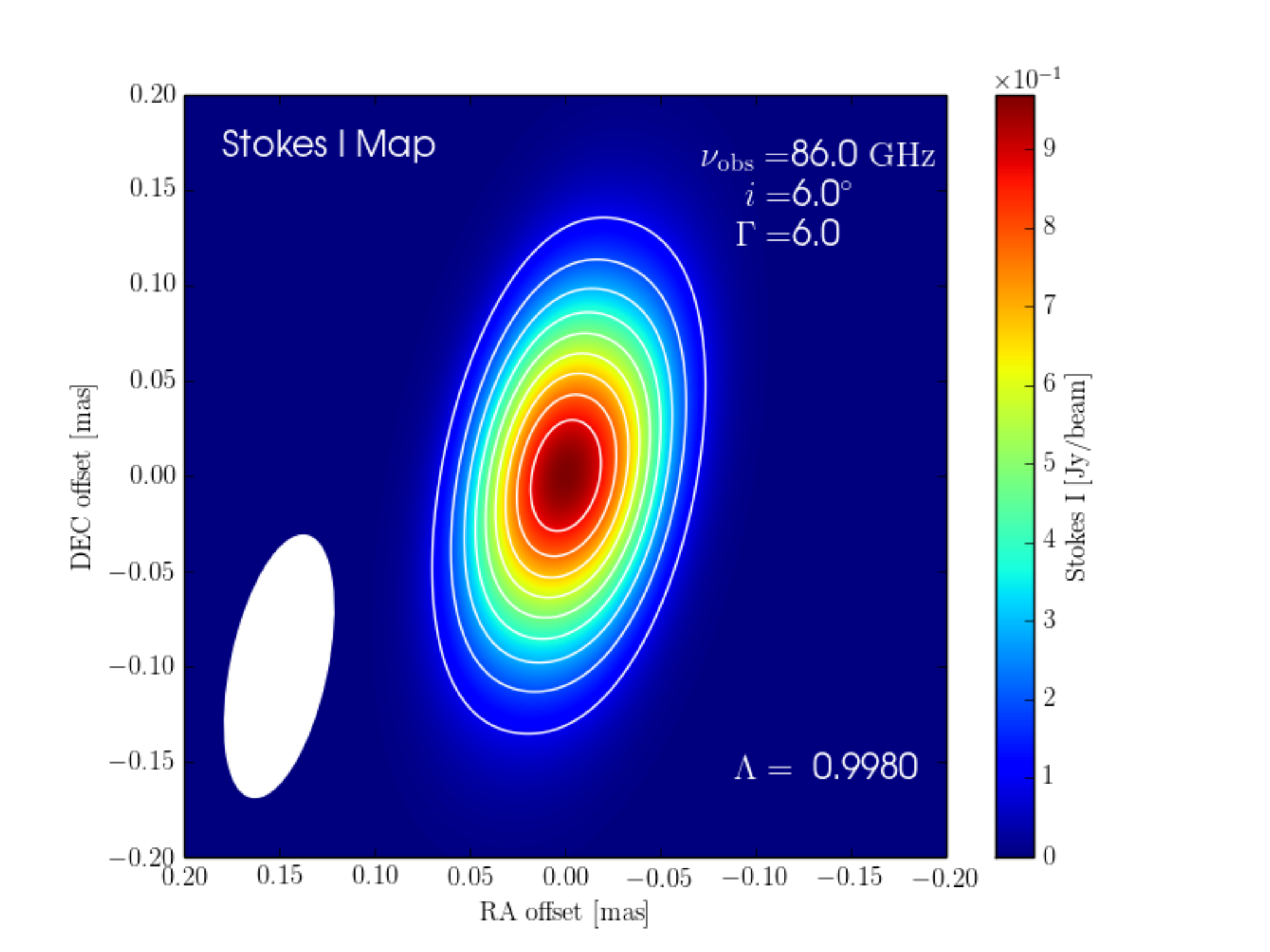}
\includegraphics[width=0.49\linewidth]{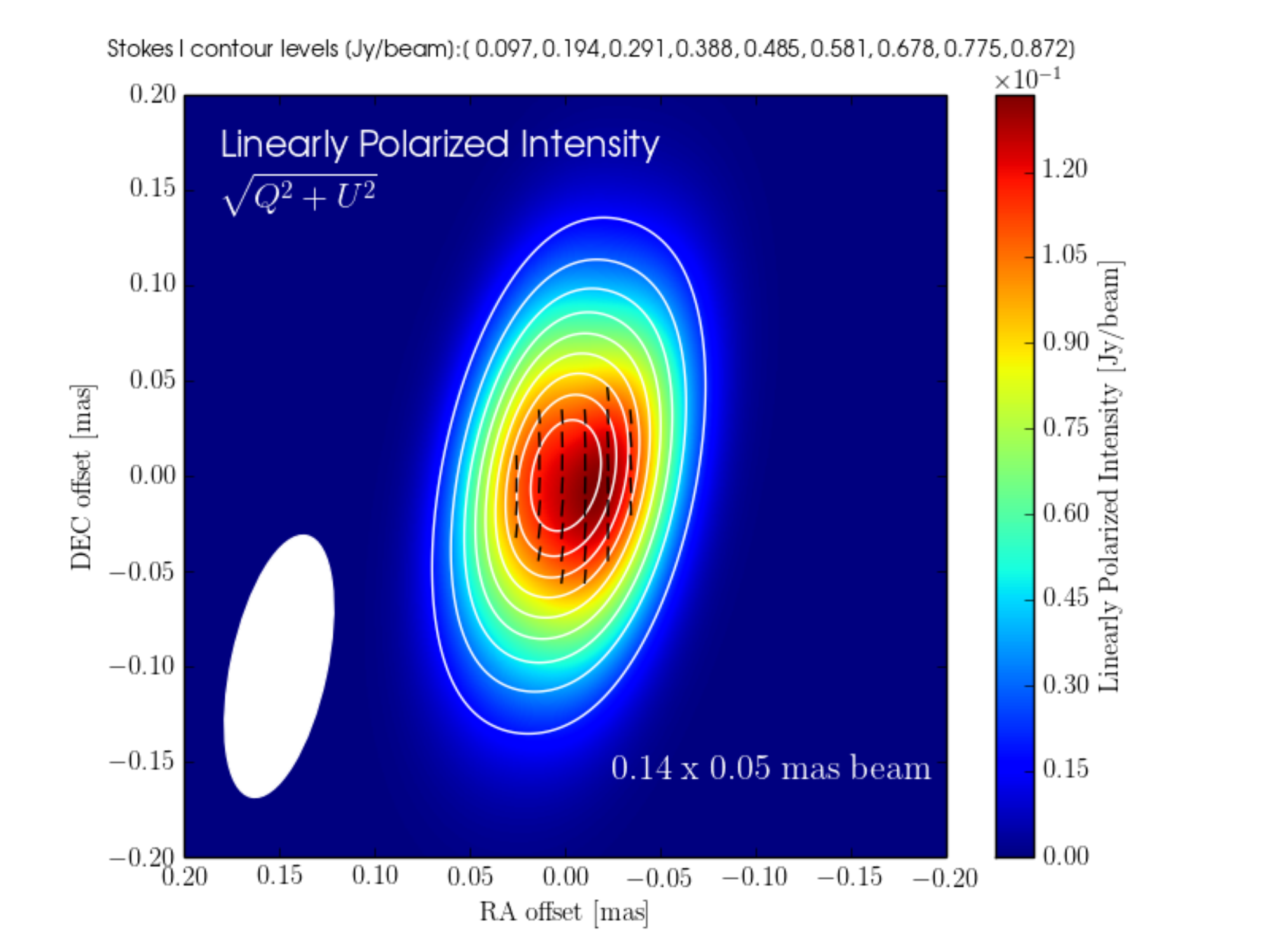}
\includegraphics[width=0.49\linewidth]{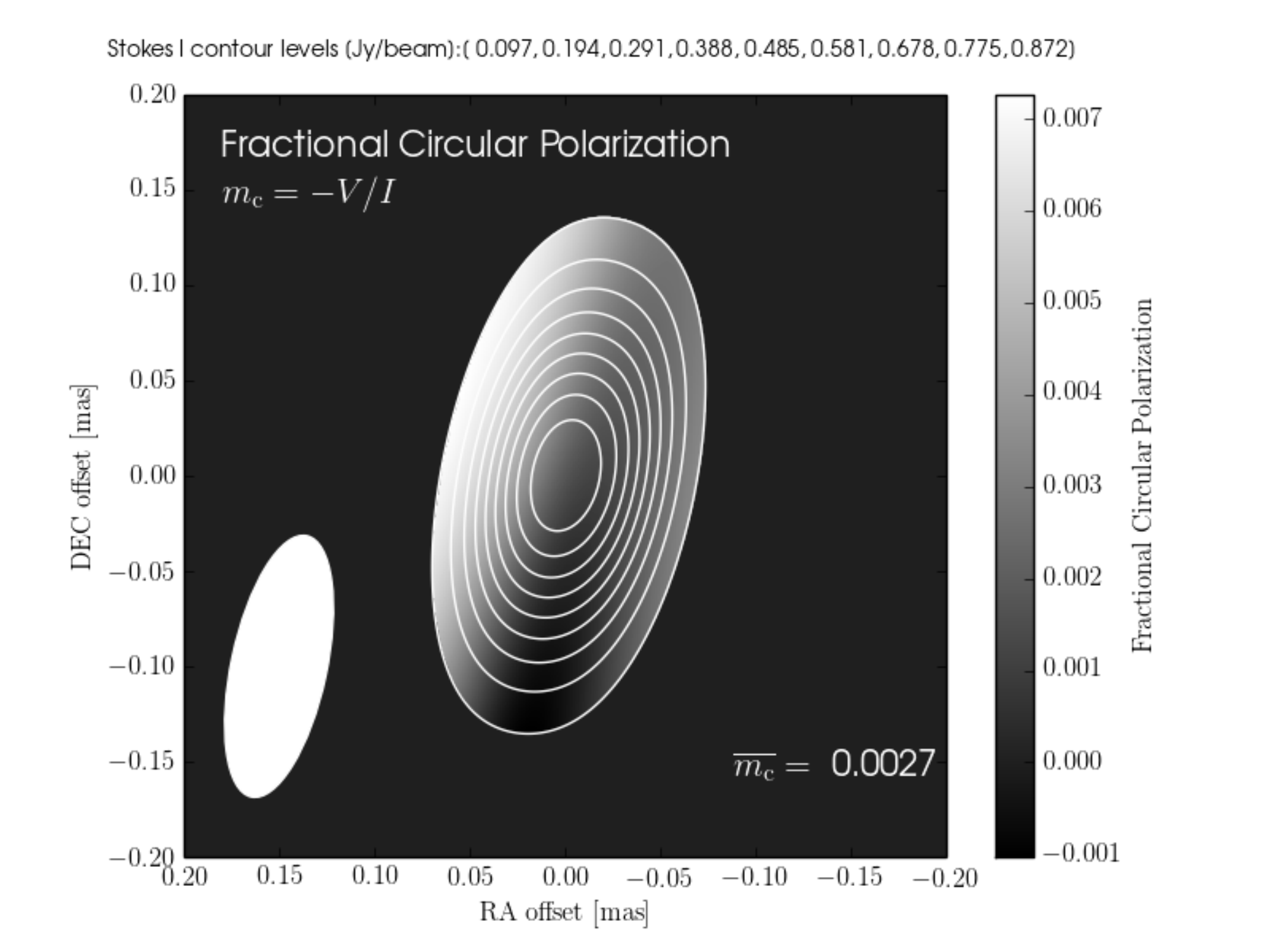}
\caption{\label{fig13} \textbf{(Upper left panel)} - A rendering of the total intensity $I$ at an observing frequency of 86 GHz. The magnetic field has the disordered structure shown in the right panel of Figure \ref{fig2}. \textbf{(Upper right panel)} - Rendering of linearly polarized intensity P of the disordered TEMZ grid.  Black line segments indicate the electric vector position angles (EVPAs) as projected onto the plane of the sky (I contours are overlaid in white). The effects of relativistic aberration (see \citealt{lyutikov05}) on the orientation of these EVPAs have been included in these calculations. \textbf{(Lower panel)} - A plot of the fractional CP in the observer's frame highlighting the different regions within the jet that produce positive and negative CP. An integrated value (see discussion in \S7) of CP is listed to the lower right. The above images have all been convolved with the elliptical Gaussian beam of FWHM $0.14 \times 0.05 ~\rm{mas}$ shown in the lower left of each panel. $\Lambda$ (listed in the upper left panel) pertains to the plasma composition of the jet and is discussed in \S10.}
\end{figure*}

\noindent As discussed in \S6, the majority of the synthetic images presented in this paper are of scales within the jet that are smaller than the maximum resolution currently attainable with ground based interferometric arrays.  In the interest of making direct comparisons to current (and upcoming) mm-wave VLBI blazar observations, we present in this appendix a series of images identical to those presented in Figure \ref{fig5}, but that have instead been convolved with a elliptical Gaussian beam of FWHM~$0.14 \times 0.05$ mas.  These images (Figure \ref{fig13}) mimic the resolution attainable with the longest baselines available to the Global Millimeter VLBI Array (GMVA).  

\end{document}